\documentclass{IEEEtran}
\usepackage{xcolor,soul,framed} 

\colorlet{shadecolor}{yellow}
\usepackage[pdftex]{graphicx}
\graphicspath{{../pdf/}{../jpeg/}}
\DeclareGraphicsExtensions{.pdf,.jpeg,.png}

\usepackage[cmex10]{amsmath}
\usepackage{array}
\usepackage{mdwmath}
\usepackage{mdwtab}
\usepackage{eqparbox}
\usepackage{url}
\usepackage{amsmath,amssymb,amsfonts}
\usepackage{cite}
\usepackage{pdfpages}
\usepackage{subfigure}
\usepackage{tabularx}
\usepackage{cuted}
\usepackage{diagbox}
\usepackage{array}  
\usepackage[numbers,sort&compress]{natbib}
\usepackage{stfloats}
\usepackage{bm}
\usepackage{upgreek}
\usepackage{comment}

\def\BibTeX{{\rm B\kern-.05em{\sc i\kern-.025em b}\kern-.08em
    T\kern-.1667em\lower.7ex\hbox{E}\kern-.125emX}}
\begin{document}
\title{
A Unified Deterministic Channel Model for Multi-Type RIS with Reflective, Transmissive, and Polarization Operations
}
\author{Yuxiang Zhang,
      Jianhua Zhang,
      Zhengfu Zhou,
      Huiwen Gong,
      Hongbo Xing,
      Zhiqiang Yuan,
      Lei Tian,
      Li Yu,
      Guangyi Liu,
     and Tao Jiang     
\thanks{Manuscript received xxx; revised xxx; accepted xxx. Date of publication xxx; date of current version xxx. This research is supported by National Key R\&D Program of China (2023YFB2904803), Young Scientists Fund of the National Natural Science Foundation of China (62201087, 62401084),  Guangdong Major Project of Basic and Applied Basic Research (2023B0303000001), Beijing Natural Science Foundation (L243002), National Natural Science Foundation of China (62341128) and Beijing University ot Posts and Telecommunications-China Mobile Research Institute Joint innovation Center.}
\thanks{Yuxiang Zhang, Jianhua Zhang, Zhengfu Zhou, Huiwen Gong, Hongbo Xing, Lei Tian and Li Yu are with the State Key Laboratory of Networking and Switching Technology, Beijing University of Posts and Telecommunications, Beijing 100876, China (e-mail: zhangyx@bupt.edu.cn; jhzhang@bupt.edu.cn; zhengfu@bupt.edu.cn; birdsplan@bupt.edu.cn; hbxing@bupt.edu.cn; tianlbupt@bupt.edu.cn; li.yu@bupt.edu.cn).}
\thanks{Zhiqiang Yuan is with the National Mobile Communications Research Laboratory, School of Information Science and Engineering, Southeast University, Nanjing 210096, China (email: zqyuan@seu.edu.cn)}
\thanks{Guangyi Liu and Tao Jiang are with the Future Research Laboratory, China Mobile Research Institute, Beijing 100053, China (e-mail: liuguangyi@chinamobile.com; jiangtao@chinamobile.com). }
}

\maketitle

\begin{abstract}
Reconfigurable Intelligent Surface (RIS) technologies have been considered as a promising enabler for 6G, enabling advantageous control of electromagnetic (EM) propagation. 
RIS can be categorized into multiple types based on their reflective/transmissive modes and polarization control capabilities, all of which are expected to be widely deployed in practical environments. A reliable RIS channel model is essential for the design and development of RIS communication systems. While deterministic modeling approaches such as ray-tracing (RT) offer significant benefits, a unified model that accommodates all RIS types is still lacking.
This paper addresses this gap by developing a high-precision deterministic channel model based on RT, supporting multiple RIS types: reflective, transmissive, hybrid, and three polarization operation modes.
To achieve this, a unified EM response model for the aforementioned RIS types is developed. The reflection and transmission coefficients of RIS elements are derived using a tensor-based equivalent impedance approach, followed by calculating the scattered fields of the RIS to establish an EM response model.
The performance of different RIS types is compared through simulations in typical scenarios. During this process, passive and lossless constraints on the reflection and transmission coefficients are incorporated to ensure fairness in the performance evaluation.
Simulation results validate the framework's accuracy in characterizing the RIS channel, and specific cases tailored for dual-polarization independent control and polarization rotating RISs are highlighted as insights for their future deployment. This work can be helpful for the evaluation and optimization of RIS-enabled wireless communication systems.

\end{abstract}

\begin{IEEEkeywords}
RIS, Deterministic channel model, Ray-tracing, EM response model, Polarization control
\end{IEEEkeywords}

\section{Introduction}
\IEEEPARstart{R}{econfigurable} Intelligent Surface (RIS) is a two-dimensional surface composed of numerous elements, where the reflection/transmission coefficient of each element can be actively adjusted \cite{smartRadio,cui,zhang}. By dynamically adjusting the phase and amplitude of reflection and transmission coefficients at each element, RIS enables unprecedented control over the wireless environment, such as enhancing coverage \cite{TVT_RISperfermance,RIS_UAV_perfermance}, improving channel rank conditions \cite{zhangrui}, and suppressing interference \cite{jinshi_TVT_RISchannelmodeling}. In March 2025, 3GPP plans to hold a TSG-wide 6G workshop to initiate discussions on potential technologies for future releases, and RIS is anticipated to be a promising technology considered during these discussions.

An accurate channel model is a prerequisite for RIS system design \cite{3dmimo,twc_RISmeasurement}, as it provides an indispensable platform to evaluate the performance of RIS technologies and optimize their deployment. 
Ray-tracing (RT) has been widely adopted as a deterministic channel modeling technique, capable of simulating the signal propagation process with high precision \cite{yuan_TCT_RT,He,yuan2023sub,RT_GBSM_Analysis}.
It serves as a powerful tool for channel simulation \cite{yuan}, making it a natural candidate for RIS channel modeling.
However, most traditional RT methods fail to incorporate RIS in channel modeling and lack the capability to support scenarios involving multiple types of RIS. To develop an accurate RT-based channel model for RIS, it is essential to precisely characterize the electromagnetic (EM) response of RIS.

Current studies on RIS EM response modeling primarily focus on simple types, such as polarization-coherent reflective and transmissive RIS \cite{RT_RIS_reflect}. However, such models fail to comprehensively capture the capabilities of all RIS types. For example, categorized by operational modes, RIS can be classified into reflective, transmissive \cite{transmission}, or reflective-transmissive types \cite{Yuanwei}, and by polarization control capabilities as dual-polarization unified control, dual-polarization independent control \cite{dual_RIS1,Polar_rotation}, and polarization-rotating RIS \cite{alltypeRIS,rotation_RIS}. Ignoring the operational degrees of freedom of these RIS types will limit the functionality and applicability of RIS. However, to the best of the authors' knowledge, studies addressing EM modeling for RIS with complex polarization operations remain limited.

Specifically, early research on RIS EM response models focuses on simple reflective or transmissive types. For instance, polarization-coherent RIS are analyzed using physical optics and the equivalence principle to calculate reflection coefficients and secondary radiation \cite{RIS_PO1, RIS_PO2}.
In addition to the reflection and transmission types, research on RIS has also explored polarization control capabilities. Models for dual-polarized independent control RIS have highlighted the importance of incorporating angle dependence into EM response calculations \cite{dou}. Research on polarization-rotating RIS, though limited, includes studies proposing novel metamaterial designs for dual-direction polarization conversion \cite{alltypeRIS} or electrically controlled polarization rotation \cite{rotation_RIS}.
A further dimension of research investigates the feasibility of achieving arbitrary reflection and transmission coefficient modulation. While early studies assumed that RIS could realize any desired coefficient, it has since been shown that the achievable coefficients are inherently angle-dependent \cite{zhengfu, ghw, electromagneticRIS, Tang1}.

Despite these advancements, no existing model fully integrates these three dimensions. This gap becomes particularly problematic for multi-polarization scenarios, where complex interactions between polarization, reflection/transmission, and angle coupling require additional constraints, such as passive and lossless conditions. Furthermore, current models often overlook the dependence of RIS reflection/transmission coefficients on incident angles, a critical factor for accurate channel modeling. 

To address these challenges, this paper proposes a unified and deterministic RIS channel modeling framework that supports diverse RIS types and ensures consistent performance evaluation, the contributions of this paper are summarized as follows:
(i) We propose a deterministic RT RIS channel model capable of characterizing all known types of RIS, encompassing reflective, transmissive, and dual-polarization unified control, dual-polarization independent control, and polarization-rotating operation configurations.
Meanwhile, angle-dependent reflection and transmission coefficients are derived for all these types, enabling more accurate modeling.
(ii) Passive and lossless conditions are considered to ensure fair performance comparisons among different RIS types in identical scenarios. Simulation results validate the framework’s accuracy in characterizing the RIS channel.

The rest of the paper is organized as follows: Section II provides the overall expression of the RIS channel. Section III details the establishment of the EM response for RIS elements and presents the passive and lossless conditions for the three types of RIS elements. In Section IV, the impact of incident angle dependence on the performance of RIS is analyzed, and a fair comparison of the performance of the three types of RIS is conducted. Section V concludes the paper.

Notations: For simplicity, dual-polarization unified control, dual-polarization independent control, and polarization-rotating RIS are referred to as \textit{Type 1/2/3 RIS} in the below. Fonts $a$, ${\mathrm{a}}$, and $\boldsymbol{\mathrm{A}}$ represent scalars, vectors, and matrices, respectively; j corresponds to the imaginary unit; $(\cdot)^H$, $(\cdot)^T$, and $(\cdot)^*$ represent Hermitian, transpose, and conjugate, respectively; $\overline{\overline{\mathbf{A}}}$ denotes a tensor; $\nabla$ denotes the nabla operator; $\cdot$ and $\times$ denotes the inner product and cross product, respectively; $\overline{\overline{\mathbf{I}}}$ denote the identity matrix with compatible dimensions; $\|\cdot\|$ denotes denotes $\ell2$-norm of the vector; $\mathbf{\hat{x}}$, $\mathbf{\hat{y}}$ and $\mathbf{\hat{z}}$ denote the unit vector of $x$-direction, $y$-direction and $z$-direction in three-dimensional Cartesian coordinate coordinate system, respectively; $\mathbf{\hat{\theta}}$ and $\mathbf{\hat{\phi}}$ denote the unit vector of $\theta$-direction and $\phi$-direction in spherical coordinate system, respectively.
In particular, $\boldsymbol{\mathrm{E}}=\left(E_x,E_y,E_z\right)^T\in \mathbb{C}^{3\times1}$ denotes the electric field while $\boldsymbol{\mathrm{H}}=\left(H_x,H_y,H_z\right)^T\in \mathbb{C}^{3\times1}$ denotes the magnetic field.

\section{system model}
\begin{figure*}
  \begin{center}
  \includegraphics[width=0.8\linewidth]{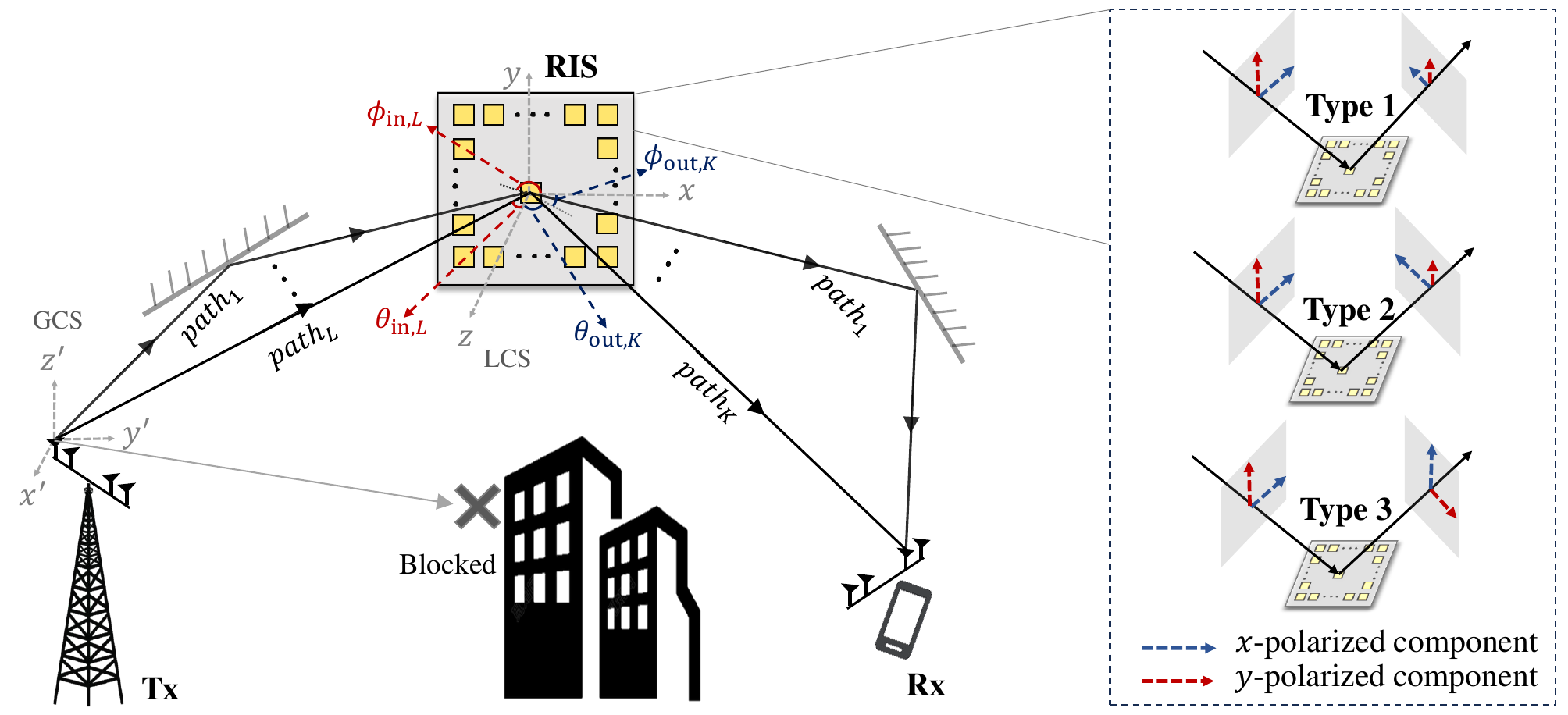}
  \caption{Illustration of RIS-assisted communication channels under three types of RIS configurations.}\label{RIS_channel}
  \end{center}
\end{figure*}
In this section, we present the expression for the RIS channel, which will underpin the subsequent work.

Fig. \ref{RIS_channel} illustrates the RIS channel between the $u$-th transmit antenna and the $s$-th receive antenna. It is clear that this channel primarily consists of the Tx-Rx direct channel and the Tx-RIS-Rx cascaded channel. The figure also shows the contribution of the $n$-th RIS element to the Tx-RIS-Rx cascaded channel. In reality, due to the multitude of RIS elements on the RIS panel, the Tx-RIS-Rx cascaded channel is the aggregate of contributions from each RIS element. Thus, the RIS channel can be accurately described as follows:
\begin{equation}
\begin{aligned}
   h_{u,s}^{pq}(\tau) &= h^{D,pq}_{u,s}(\tau) + h^{C,pq}_{u,s}(\tau) \\
   &= h^{D,pq}_{u,s}(\tau) + \sum_{n=1}^{N} h^{C,pq}_{u,s,n}(\tau),
   \label{fundamental_diode_voltage}
\end{aligned}
\end{equation}
where $h^{D}$ represents the Tx-Rx direct channel, $h^{C}$ represents the Tx-RIS-Rx cascaded channel, $N$ is the number of RIS elements, and $h^{C}_{n}$ is the contribution of the $n$-th RIS element. The superscript $pq$ means transmitting with polarization $q$ and receiving with polarization $p$. Here, $p,q\in \{\mathbf{\hat{\theta}}',\mathbf{\hat{\phi}}'\}$, where $\mathbf{\hat{\theta}}'$ and $\mathbf{\hat{\phi}}'$ represent $\mathbf{\hat{\theta}}'$-polarization and $\mathbf{\hat{\phi}}'$-polarization in global coordinate system, respectively (see \cite{3gpp}). Furthermore, $h^{C}_{n}$ consists of the following three components:

Firstly, the EM response of $n$-th RIS element characterizes the changes in amplitude, phase, and polarization of the signal departing from the RIS element compared to the incident signal, which can be represented as 
\begin{equation}\label{eq2}
\mathbf{G}'(\theta'_{\mathrm{in}}, \phi'_{\mathrm{in}}, \theta'_{\mathrm{out}}, \phi'_{\mathrm{out}}, \overline{\overline{\mathbf{R}}}, \overline{\overline{\mathbf{T}}}) = \\
\begin{bmatrix}
    G'^{\mathbf{\hat{\theta}}'\mathbf{\hat{\theta}}'}& G'^{\mathbf{\hat{\theta}}'\mathbf{\hat{\phi}}'}\\
     G'^{\mathbf{\hat{\phi}}'\mathbf{\hat{\theta}}'}& G'^{\mathbf{\hat{\phi}}'\mathbf{\hat{\phi}}'}
\end{bmatrix},
\end{equation}
where $\theta'_{\mathrm{in}}$ and $\phi'_{\mathrm{in}}$ represent the zenith and azimuth angle of the incident path, $\theta'_{\mathrm{out}}$ and $\phi'_{\mathrm{out}}$ represent the zenith and azimuth angle of departure path, respectively. $\overline{\overline{\mathbf{R}}}$ and $\overline{\overline{\mathbf{T}}}$ are predefined desired reflection and transmission coefficients. $G'^{pq}$  represents the conversion from the $q$-polarized incident signals to $p$-polarized depature signals. 

Since the local coordinate system of the RIS does not necessarily align with the global coordinate system, the EM response of the RIS elements can be obtained through a transformation from the EM response under the local coordinate system, which can be represented as
\begin{equation}\label{eq3}
\begin{split}
\mathbf{G}'(&\theta'_{\mathrm{in}}, \phi'_{\mathrm{in}}, \theta'_{\mathrm{out}}, \phi'_{\mathrm{out}}, \overline{\overline{\mathbf{R}}}, \overline{\overline{\mathbf{T}}}) \\=&
\mathbf{P}\cdot \mathbf{G}(\theta_{\mathrm{in}}, \phi_{\mathrm{in}}, \theta_{\mathrm{out}}, \phi_{\mathrm{out}}, \overline{\overline{\mathbf{R}}}, \overline{\overline{\mathbf{T}}}) \cdot \mathbf{P}^{-1}
\\=&\mathbf{P}\cdot 
\begin{bmatrix}
    G^{\mathbf{\hat{\theta}}\mathbf{\hat{\theta}}}& G^{\mathbf{\hat{\theta}}\mathbf{\hat{\phi}}}\\
     G^{\mathbf{\hat{\phi}}\mathbf{\hat{\theta}}}& G^{\mathbf{\hat{\phi}}\mathbf{\hat{\phi}}}
\end{bmatrix}
\cdot \mathbf{P}^{-1}
,
\end{split}
\end{equation}
where $ \mathbf{G}(\theta_{\mathrm{in}}, \phi_{\mathrm{in}}, \theta_{\mathrm{out}}, \phi_{\mathrm{out}}, \overline{\overline{\mathbf{R}}}, \overline{\overline{\mathbf{T}}})$ is the RIS element EM response under the local coordinate system, $\theta_{\mathrm{in}}$, $\phi_{\mathrm{in}}$, $\theta_{\mathrm{out}}$ and $\phi_{\mathrm{out}}$ are the angles under the local coordinate system, $\mathbf{\hat{\theta}}$ and $\mathbf{\hat{\phi}}$ represent $\mathbf{\hat{\theta}}$ and $\mathbf{\hat{\phi}}$ polarization under local coordinate system, and $\mathbf{P}$ represents the transformation matrix from $\mathbf{\hat{\theta}}$ and $\mathbf{\hat{\phi}}$ polarization to $\mathbf{\hat{\theta}}'$ and $\mathbf{\hat{\phi}}'$ polarization, as detailed in (7.1-9) in \cite{3gpp}.

Secondly, the excitation of the EM response of $n$-th RIS element originates from the channel between Tx and $n$-th RIS element, which can be modeled as
\begin{equation}\label{fundamental_diode_voltage}
   h_{u,s,n}^{Tx-RIS,pq}(\tau)=\sum_{l=1}^{L} \alpha_{n,l}^{pq}e^{j\gamma_{n,l}^{pq}} \delta(\tau - \tau_{l}),
\end{equation}
where $L$ represents the number of multipaths in the Tx-RIS channel, $\alpha_{n,l}$, $\gamma_{n,l}$ and $\tau_{l}$ represent the amplitude, phase and delay of the $l$-th multipath. All of these subchannel data can be provided by RT.

Finally, the signal departing from the $n$-th RIS element will propagate through the channel between the $n$-th RIS element and Rx, which can be modeled as
\begin{equation}\label{fundamental_diode_voltage}
   h_{u,s,n}^{RIS-Rx,pq}(\tau)=\sum_{k=1}^{K} \alpha_{n,k}^{pq}e^{j\gamma_{n,k}^{pq}} \delta(\tau - \tau_{k}),
\end{equation}
where $K$ represents the number of multipaths in the RIS-Rx channel, $\alpha_{n,k}$, $\gamma_{n,k}$ and $\tau_{k}$ represent the amplitude, phase and delay of the $k$-th multipath. All of these subchannel data can be produced by RT.

Taking into account the above three processes, the contribution of the $n$-th RIS element can be expressed as
\begin{equation}\label{eq5}
\begin{split}
   h_{u,s,n}^{C,pq}(\tau) = &\sum_{l=1}^{L} \sum_{k=1}^{K} 
\begin{bmatrix}
    \alpha_{n,k}^{p\mathbf{\hat{\theta}}'}e^{j\gamma_{n,k}^{p\mathbf{\hat{\theta}}'}}\\
     \alpha_{n,k}^{p\phi'}e^{j\gamma_{n,k}^{p\phi'}}
\end{bmatrix}^{T}\\
&\cdot \mathbf{G}'(\theta'_{\mathrm{in},l}, \phi'_{\mathrm{in},l}, \theta'_{\mathrm{out},k}, \phi'_{\mathrm{out},k}, \overline{\overline{\mathbf{R}}}, \overline{\overline{\mathbf{T}}})\\ 
&\cdot 
\begin{bmatrix}
   \alpha_{n,l}^{\mathbf{\hat{\theta}}' q}e^{j\gamma_{n,l}^{\mathbf{\hat{\theta}}' q}}\\
    \alpha_{n,l}^{\mathbf{\hat{\phi}}' q}e^{j\gamma_{n,l}^{\mathbf{\hat{\phi}}' q}}
\end{bmatrix}
\cdot \delta(\tau - \tau_{l}-\tau_{k}).
\end{split}
\end{equation}
It can be observed that the number of multipaths is $LK$, this is because each multipath in the $h_{n}^{Tx-RIS}$ can excite $K$ multipaths after reaching the $n$-th RIS element.

\section{RIS electromagnetic response modeling}
In this section, the EM response model of the RIS elements is established. According to \eqref{eq3}, it is sufficient to calculate only $\mathbf{G}(\theta_{\mathrm{in}}, \phi_{\mathrm{in}}, \theta_{\mathrm{out}}, \phi_{\mathrm{out}}, \overline{\overline{\mathbf{R}}}, \overline{\overline{\mathbf{T}}})$.

Specifically, we first introduce the impedance-type generalized sheet transition conditions (GSTCs) hardware model in subsection A. This model treats the RIS element as a sheet with specific impedance tensors and provides boundary conditions that constrain the EM fields on both sides of the RIS element. It can depict the incident angle dependence of the RIS element's reflection and transmission coefficients. In subsection B, the actual reflection/transmission coefficients of the three types of RIS elements at a given incident angle are calculated based on the impedance-type GSTCs model. Thus, the EM field distribution on the RIS element's surface can be determined. In subsection C, we then consider the RIS element surface with its EM field distribution as a secondary radiation source and calculate the RIS EM response. In subsection D, the passive and lossless conditions are presented, which can be used to fairly compare the performance of different types of RIS.
\subsection {Impedance-type GSTCs model}
\begin{figure}
  \begin{center}
  \includegraphics[width=3.1in]{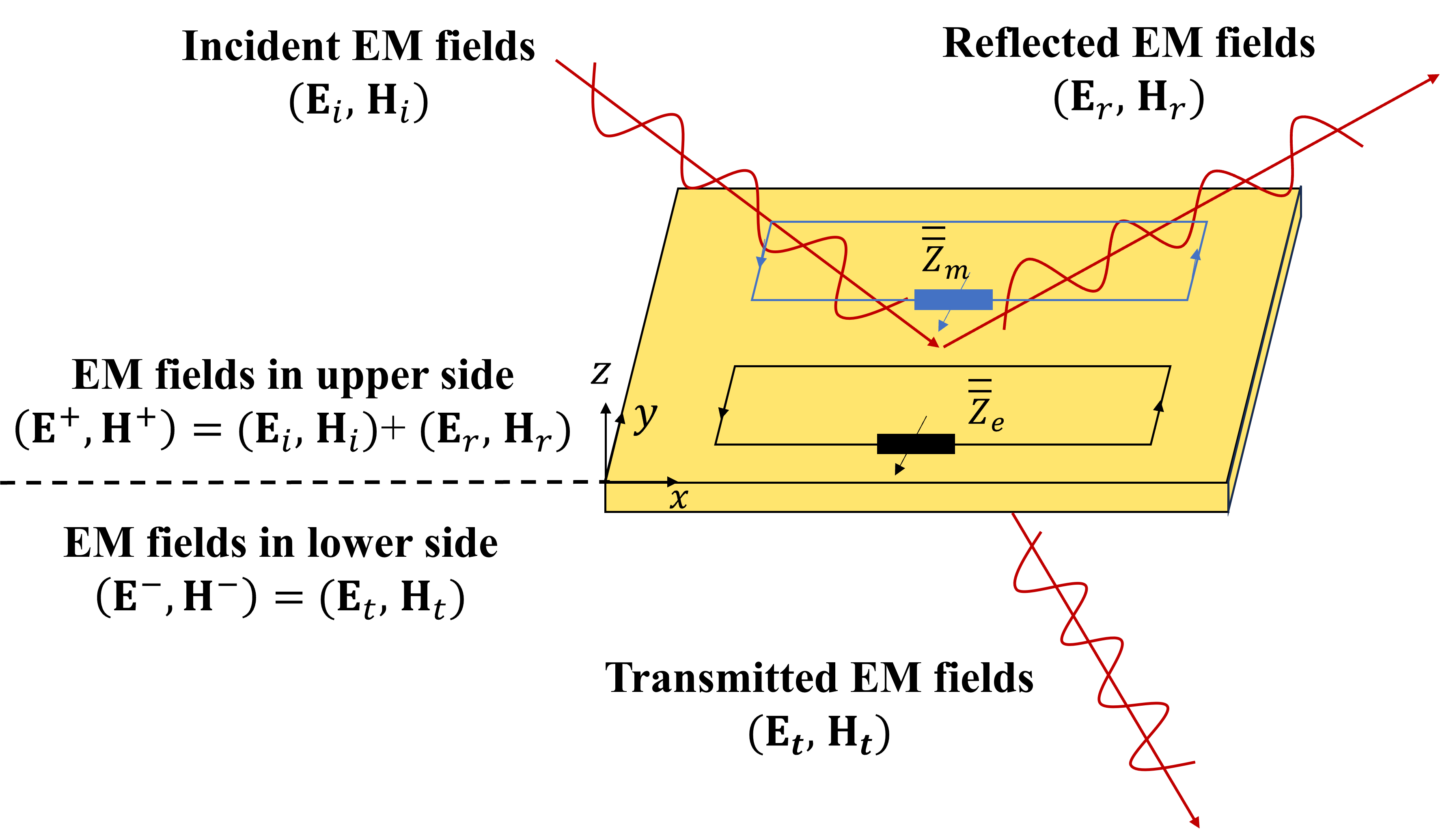}
  \caption{Impedence-type GSTCs EM model of RIS element}\label{RIS_element}
  \end{center}
\end{figure}
As depicted in Fig. \ref{RIS_element}, consider a RIS element in free space, and establish a local coordinate system. In this context, the impedance-type GSTCs \cite{surfaceEM} can be formulated as
\begin{subequations}\label{eqn-6}
  \begin{align}
       \mathbf{\hat{z}} \times (\mathbf{E}^{+} - \mathbf{E}^{-}) &= -\overline{\overline{\mathbf{Z}}}_{m}\mathbf{H}_{t,av},\\
     \mathbf{\hat{z}} \times (\mathbf{H}^{+} - \mathbf{H}^{-}) &= \overline{\overline{\mathbf{Y}}}_{e} \mathbf{E}_{t,av},
  \end{align}
  \label{eq6}
\end{subequations}
where $\overline{\overline{\mathbf{Y}}}_{e}$, $\overline{\overline{\mathbf{Z}}}_{m}\in \mathbb{C} ^{3 \times 3} $ represent the electrical admittance and magnetic impedance tensors of the RIS element, collectively referred to as impedance tensors. $\mathbf{E}^{+}$ and $\mathbf{H}^{+}$ stand for the EM fields on the upper side, which are equal to the superposition of the incident and reflected fields. Meanwhile, $\mathbf{E}^{-}$ and $\mathbf{H}^{-}$ represent the EM fields on the lower side, which are determined by the transmitted fields. $\mathbf{E}_{t,av}$ and $\mathbf{H}_{t,av}$ represent the tangential components of the average fields $\mathbf{E}_{av}$ and $\mathbf{H}_{av}$ on both sides, respectively, and have
\begin{subequations}\label{eqn-7}
  \begin{align}
\mathbf{E}_{av} &= \frac{\mathbf{E}^{+}+\mathbf{E}^{-}}{2},\\
\mathbf{H}_{av} &= \frac{\mathbf{H}^{+}+\mathbf{H}^{-}}{2},\\
\mathbf{E}_{t,av}&=\mathbf{E}_{av}-\left(\mathbf{E}_{av}\cdot\mathbf{\hat{z}}\right)\mathbf{\hat{z}},\\
\mathbf{H}_{t,av}&=\mathbf{H}_{av}-\left(\mathbf{H}_{av}\cdot\mathbf{\hat{z}}\right)\mathbf{\hat{z}}.
  \end{align}
\end{subequations}

In the impedance tensors, there exists redundancy in the elements, as mentioned in \cite{GTSC1}. Consequently, for Type 1/2/3 RIS elements, the impedance tensors can be simplified to
\begin{subequations}\label{eqn-8}
  \begin{align}
\overline{\overline{\mathbf{Y}}}_{e} &= 
\begin{bmatrix}
Y_{xx} & Y_{xy} & 0 \\
Y_{yx} & Y_{yy} & 0 \\
0 & 0 & 0
\end{bmatrix}
,\\
\overline{\overline{\mathbf{Z}}}_{m} &= 
\begin{bmatrix}
Z_{xx} & Z_{xy} & 0 \\
Z_{yx} & Z_{yy} & 0 \\
0 & 0 & 0
\end{bmatrix}
.
  \end{align}
\end{subequations}
Thus, the $z$-direction components of the EM fields in \eqref{eq6} are inactive, allowing for their simultaneous simplification as
\begin{subequations}
  \begin{align}\label{eqn-10a}
\begin{bmatrix}
E_{y}^{-} - E_{y}^{+} \\
E_{x}^{+} - E_{x}^{-} \\
\end{bmatrix}
&=-
\begin{bmatrix}
Z_{xx} & Z_{xy} \\
Z_{yx} & Z_{yy} \\
\end{bmatrix}
\begin{bmatrix}
H_{av,x} \\
H_{av,y} \\
\end{bmatrix},\\
\label{eqn-10b}
\begin{bmatrix}
H_{y}^{-} - H_{y}^{+} \\
H_{x}^{+} - H_{x}^{-} \\
\end{bmatrix}
&=
\begin{bmatrix}
Y_{xx} & Y_{xy} \\
Y_{yx} & Y_{yy} \\
\end{bmatrix}
\begin{bmatrix}
E_{av,x} \\
E_{av,y} \\
\end{bmatrix}.
  \end{align}
  \label{eq9}
\end{subequations}

Equations (\ref{eqn-10a}) and (\ref{eqn-10b}) will play a pivotal role in this paper, serving as the foundation for subsequent analysis of the RIS element reflection/transmission coefficients. 

Additionally, if the RIS element is assumed to be passive and lossless \cite{GTSC2}, its impedance tensors must adhere to the following conditions:
\begin{equation}\label{eq10}
    \overline{\overline{\mathbf{Y}}}_{e}^{*} = \overline{\overline{\mathbf{Y}}}_{e}^{T}, \quad \overline{\overline{\mathbf{Z}}}_{m}^{*} = \overline{\overline{\mathbf{Z}}}_{m}^{T}.
\end{equation}

\subsection{Incident angle dependence of reflection/transmission coefficients}
The reflection and transmission coefficients $\overline{\overline{\mathbf{R}}}$ and $\overline{\overline{\mathbf{T}}}$ of each RIS element can be actively configured by the user to achieve various functions, such as beamforming and beamfocusing. However, since a significant number of RIS elements are typically designed for normal incident signals \cite{GTSC1,GTSC2,normalRIS}, the actual reflection and transmission coefficients may deviate from the predefined desired values when the signal is obliquely incident, exhibiting incident angle dependence \cite{ghw,electromagneticRIS,Tang1}. Therefore, to accurately characterize the EM response of RIS elements, it is crucial to establish a mapping relationship between the actual reflection/transmission coefficients and the predefined desired reflection/transmission coefficients. 

The predefined desired reflection and transmission coefficients of a RIS element can be represented as 
\begin{equation}\label{fundamental_diode_voltage}
    \overline{\overline{\mathbf{R}}} = 
\begin{bmatrix}
R_{xx} & R_{xy} \\
R_{yx} & R_{yy}
\end{bmatrix}
,
\end{equation}
and
\begin{equation}\label{fundamental_diode_voltage}
    \overline{\overline{\mathbf{T}}} = 
\begin{bmatrix}
T_{xx} & T_{xy} \\
T_{yx} & T_{yy}
\end{bmatrix}
,
\end{equation}
where $R_{ab}$ and $T_{ab}$ represent the reflection and transmission coefficients that transform the $b$-polarized incident wave to $a$-polarized depature wave. Assuming $\mathbf{\hat{\phi}}$-polarized signals incident at angles $(\theta_{\mathrm{in}}, \phi_{\mathrm{in}})$, we below illustrate the process to calculate the actual reflection/transmission coefficients.

\subsubsection{Dual-polarized unified control (Type 1) RIS element} For normal incident signals, a Type 1 RIS element does not distinguish the polarization direction and perform unified control on all signal components. Thus, only a set of uniform reflection and transmission coefficients can be configured by users, represented as $R$ and $T$. Hence, its predefined desired reflection and transmission coefficient tensors are modeled as
\begin{equation}
 \overline{\overline{\mathbf{R}}}
=
\begin{bmatrix}
R & 0 \\
0 & R
\end{bmatrix}
, \quad
 \overline{\overline{\mathbf{T}}}
=
\begin{bmatrix}
T & 0 \\
0 & T
\end{bmatrix}
.
\end{equation}

Then, the EM fields in \eqref{eq9} can be reformulated using predefined desired reflection/transmission coefficients. If the $x$-polarized EM fields $(E_{x}^{i}\mathbf{\hat{x}},H_{y}^{i}\mathbf{\hat{y})}$ normally incident on the RIS element, there is
\begin{equation}\label{fundamental_diode_voltage}
\begin{bmatrix}
E_{x}^{+} & E_{y}^{+} \\
E_{x}^{-} & E_{y}^{-} \\
H_{x}^{+} & H_{y}^{+} \\
H_{x}^{-} & H_{y}^{-}
\end{bmatrix}
=
\begin{bmatrix}
(1+R)E^{i}_{x} & 0 \\
TE^{i}_{x} & 0 \\
0 & (1-R)H^{i}_{y} \\
0 & TH^{i}_{y}
\end{bmatrix}.
\end{equation}
If the $y$-polarized EM fields $(E_{y}^{i}\mathbf{\hat{y}},H_{x}^{i}\mathbf{\hat{x}})$ normally incident on the RIS element, there is
\begin{equation}\label{fundamental_diode_voltage}
\begin{bmatrix}
E_{x}^{+} & E_{y}^{+} \\
E_{x}^{-} & E_{y}^{-} \\
H_{x}^{+} & H_{y}^{+} \\
H_{x}^{-} & H_{y}^{-}
\end{bmatrix}
=
\begin{bmatrix}
0 & (1+R)E^{i}_{y} \\
0 & TE^{i}_{y} \\
(1-R)H^{i}_{x} & 0 \\
TH^{i}_{x} & 0
\end{bmatrix}.
\end{equation}

Meanwhile, according to the properties of plane waves, the electric and magnetic fields are perpendicular to each other, the direction satisfies the right-hand rule, and the ratio of the amplitude is equal to the wave impedance, we can get
\begin{equation}\label{fundamental_diode_voltage}
E_{x}^{i} = -\eta H_{y}^{i}, \quad E_{y}^{i} = \eta H_{x}^{i},
\end{equation}
where $\eta$ is the free space wave impedance, which is $120\pi (\Omega)$.

By substituting the above equations into \eqref{eq9} and solving them, the impedance tensors of the Type 1 RIS element are determined as
\begin{subequations}\label{eq18}
  \begin{align}
\overline{\overline{\mathbf{Y}}}_{e}
=
\begin{bmatrix}
\frac{2}{\eta}\frac{1-R-T}{1+R+T} & 0 & 0\\
0 & \frac{2}{\eta}\frac{1-R-T}{1+R+T} & 0\\
0 & 0 & 0
\end{bmatrix},\\
\overline{\overline{\mathbf{Z}}}_{m}
=
\begin{bmatrix}
2\eta\frac{1+R-T}{1-R+T} & 0 & 0\\
0 & 2\eta\frac{1+R-T}{1-R+T} & 0\\
0 & 0 & 0
\end{bmatrix}.
  \end{align}
\end{subequations}

Let $R_{\phi}(\theta_{\mathrm{in}})$ and $T_{\phi}(\theta_{\mathrm{in}})$ denote the actual reflection and transmission coefficients for $\mathbf{\hat{\phi}}$-polarized signals incident at angles $(\theta_{\mathrm{in}}, \phi_{\mathrm{in}})$. The fields described in \eqref{eq9} can be re-expressed as
\begin{subequations}\label{eq27}
  \begin{align}
&E^{+}_{x} = -(1+R_{\phi}(\theta_{\mathrm{in}}))E_{i}sin\phi_{\mathrm{in}}
,\\
&E^{+}_{y} = (1+R_{\phi}(\theta_{\mathrm{in}}))E_{i}cos\phi_{\mathrm{in}},\\
&H^{+}_{x} = (1-R_{\phi}(\theta_{\mathrm{in}}))E_{i}cos\theta_{\mathrm{in}} cos\phi_{\mathrm{in}}, \\
&H^{+}_{y} = (1-R_{\phi}(\theta_{\mathrm{in}}))E_{i}cos\theta_{\mathrm{in}} sin\phi_{\mathrm{in}}, \\
&E^{-}_{x} = -T_{\phi}(\theta_{\mathrm{in}})E_{i}sin\phi_{\mathrm{in}},\\
&E^{-}_{y} = T_{\phi}(\theta_{\mathrm{in}})E_{i}cos\phi_{\mathrm{in}}, \\
&H^{-}_{x} = T_{\phi}(\theta_{\mathrm{in}})H_{i}cos\theta_{\mathrm{in}} cos\phi_{\mathrm{in}}, \\
&H^{-}_{y} = T_{\phi}(\theta_{\mathrm{in}})H_{i}cos\theta_{\mathrm{in}} sin\phi_{\mathrm{in}},
  \end{align}
\end{subequations}
where $E_{i}$ and $H_{i}$ are scalar forms of $\mathbf{E}_{i}$ and $\mathbf{H}_{i}$. Additionally, $\frac{E_{i}}{H_{i}}=\eta$.  By substituting \eqref{eq27} into \eqref{eq9} and solving the equations, the actual reflection and transmission coefficients can be determined as follows:
\begin{subequations}\label{eq28}
  \begin{align}
&R_{\phi}(\theta_{\mathrm{in}}) = \frac{-\eta Y_{xx}}{2cos\theta_{\mathrm{in}}+\eta Y_{xx}}+\frac{Z_{xx}cos\theta_{\mathrm{in}}}{Z_{xx}cos\theta_{\mathrm{in}}+2\eta}
,\\
&T_{\phi}(\theta_{\mathrm{in}}) = \frac{2cos\theta_{\mathrm{in}}}{2cos\theta_{\mathrm{in}}+\eta Y_{xx}}-\frac{Z_{xx}cos\theta_{\mathrm{in}}}{Z_{xx}cos\theta_{\mathrm{in}}+2\eta},
  \end{align}
\end{subequations}
where $ Y_{xx}$ and $Z_{xx}$ are calculated using \eqref{eq18}.

As a result, the conversion from the predefined desired reflection/transmission coefficients to the actual reflection/transmission coefficients at a given incident angle for dual-polarized unified control RIS elements is achieved.
\subsubsection{Dual-polarized independent control (Type 2) RIS element} The Type 2 RIS element is capable of independently controlling the $x$-polarized and $y$-polarized signals. Its predefined desired reflection and transmission coefficients comprise two pairs: $R_{x}$ and $T_{x}$ represent the response to the $x$-polarized signal, while $R_{y}$ and $T_{y}$ represent the response to the $y$-polarized signal. 

Thus, the predefined desired reflection and transmission coefficient tensors can be modeled as
\begin{equation}\label{fundamental_diode_voltage}
\overline{\overline{\mathbf{R}}}
=
\begin{bmatrix}
R_{x} & 0 \\
0 & R_{y} \\
\end{bmatrix}
, \quad 
\overline{\overline{\mathbf{T}}}
=
\begin{bmatrix}
T_{x} & 0 \\
0 & T_{y} \\
\end{bmatrix}.
\end{equation}
Thus, the impedance tensors can be calculated as
\begin{subequations}\label{eq21}
  \begin{align}
 \overline{\overline{\mathbf{Y}}}_{e}
=
\begin{bmatrix}
\frac{2}{\eta}\frac{1-R_{x}-T_{x}}{1+R_{x}+T_{x}} & 0 & 0\\
0 & \frac{2}{\eta}\frac{1-R_{y}-T_{y}}{1+R_{y}+T_{y}} & 0\\
0 & 0 & 0
\end{bmatrix},\\
\overline{\overline{\mathbf{Z}}}_{m}
=
\begin{bmatrix}
2\eta\frac{1+R_{y}-T_{y}}{1-R_{y}+T_{y}} & 0 & 0\\
0 & 2\eta\frac{1+R_{x}-T_{x}}{1-R_{x}+T_{x}} & 0\\
0 & 0 & 0
\end{bmatrix}.
  \end{align}
\end{subequations}

Let ${R}_{\phi,x}(\theta_{\mathrm{in}})$, ${R}_{\phi,}(\theta_{\mathrm{in}})$, ${T}_{\phi,x}(\theta_{\mathrm{in}})$ and ${T}_{\phi,y}(\theta_{\mathrm{in}})$ denote the actual reflection and transmission coefficients for $\mathbf{\hat{\phi}}$-polarized signals incident at angles $(\theta_{\mathrm{in}}, \phi_{\mathrm{in}})$. They can be derived by utilizing them to re-express the fields in \eqref{eq9} and solving \eqref{eq9}. When $\phi_{\mathrm{in}} \in \{0,\frac{\pi}{2},\pi,\frac{3\pi}{2}\}$, the results are
\begin{subequations}\label{eq31}
  \begin{align}
    {R}_{\phi,x}(\theta_{\mathrm{in}}) 
    = \frac{-2(Y_{xx}\eta^{2}-Z_{yy}cos^{2}\theta_{\mathrm{in}})}{(2cos\theta_{\mathrm{in}}+Y_{xx}\eta)(2\eta+Z_{yy}cos\theta_{\mathrm{in}})},\\
    {R}_{\phi,y}(\theta_{\mathrm{in}}) 
    = \frac{-2(Y_{yy}\eta^{2}-Z_{xx}cos^{2}\theta_{\mathrm{in}})}{(2cos\theta_{\mathrm{in}}+Y_{yy}\eta)(2\eta+Z_{xx}cos\theta_{\mathrm{in}})},\\
    {T}_{\phi,x}(\theta_{\mathrm{in}}) 
    = \frac{-\eta cos\theta_{\mathrm{in}}(Y_{xx}Z_{yy}-4)}{(2cos\theta_{\mathrm{in}}+Y_{xx}\eta)(2\eta+Z_{yy}cos\theta_{\mathrm{in}})},\\
   {T}_{\phi,y}(\theta_{\mathrm{in}}) 
   = \frac{-\eta cos\theta_{\mathrm{in}}(Y_{yy}Z_{xx}-4)}{(2cos\theta_{\mathrm{in}}+Y_{yy}\eta)(2\eta+Z_{xx}cos\theta_{\mathrm{in}})},
  \end{align}
\end{subequations}
where $ Y_{xx}$, $Y_{yy}$, $Z_{xx}$ and $Z_{yy}$ are calculated using \eqref{eq21}.

As a result, the conversion from the predefined desired reflection/transmission coefficients to the actual reflection/transmission coefficients at a given incident angle for Type 2 RIS elements is achieved.

\subsubsection{Polarization-rotating (Type 3) RIS element} Type 3 RIS elements can not only perform co-polarized reflection/transmission but also rotate the polarization direction, leading to cross-polarized reflection/transmission. Therefore, its predefined desired reflection and transmission coefficients include two pairs: $R_{co}$ and $T_{co}$ represent the co-polarization response, and $R_{cro}$ and $T_{cro}$ represent the cross-polarization response (For normally incident signals, this implies the conversion of $x/y$-polarized components to $y/x$-polarized components. For obliquely incident signals, it signifies the transformation of $\mathbf{\hat{\theta}}/\boldsymbol{\hat{\upphi}}$-polarized components to $\mathbf{\hat{\phi}}/\boldsymbol{\hat{\uptheta}}$-polarized components). 

As a result, we obtain
\begin{equation}\label{fundamental_diode_voltage}
\overline{\overline{\mathbf{R}}}
=
\begin{bmatrix}
R_{co} & -R_{cro} \\
R_{cro} & R_{co} \\
\end{bmatrix}
, \quad
\overline{\overline{\mathbf{T}}}
=
\begin{bmatrix}
T_{co} & -T_{cro} \\
T_{cro} & T_{co} \\
\end{bmatrix}
,
\end{equation}
and the impedance tensors can be calculated as
\begin{subequations}\label{eqn-18}
  \begin{align}
   \mathbf{Y}_{xx} &= \frac{2}{\eta}\frac{1-(R_{co}+T_{co})^{2}-(R_{cro}+T_{cro})^{2}}{(1+R_{co}+T_{co})^{2}+(R_{cro}+T_{cro})^{2}},\\
 \mathbf{Y}_{xy} &= \frac{4}{\eta} \frac{R_{cro}+T_{cro}}{(1+R_{co}+T_{co})^{2}+(R_{cro}+T_{cro})^{2}},\\
  \mathbf{Y}_{yy} &= \frac{2}{\eta}\frac{1-(R_{co}+T_{co})^{2}-(R_{cro}+T_{cro})^{2}}{(1+R_{co}+T_{co})^{2}+(R_{cro}+T_{cro})^{2}},\\
   \mathbf{Y}_{yx} &= \frac{-4}{\eta} \frac{R_{co}+T_{co}}{(1+R_{co}+T_{co})^{2}+(R_{cro}+T_{cro})^{2}},\\
    \mathbf{Z}_{xx} &= 2\eta\frac{1-(R_{co}-T_{co})^{2}-(R_{cro}-T_{cro})^{2}}{(1-R_{co}+T_{co})^{2}+(R_{cro}-T_{cro})^{2}},\\
    \mathbf{Z}_{xy} &= -4\eta\frac{R_{cro}-T_{cro}}{(1-R_{co}+T_{co})^{2}+(R_{cro}-T_{cro})^{2}},\\
     \mathbf{Z}_{yy} &= 2\eta\frac{1-(R_{co}-T_{co})^{2}-(R_{cro}-T_{cro})^{2}}{(1-R_{co}+T_{co})^{2}+(R_{cro}-T_{cro})^{2}},\\
     \mathbf{Z}_{yx} &= 4\eta\frac{R_{cro}-T_{cro}}{(1-R_{co}+T_{co})^{2}+(R_{cro}-T_{cro})^{2}}.
  \end{align}
\end{subequations}

 Using actual reflection/transmission coefficients of the Type 3 RIS element to re-expressed the fields in \eqref{eq9} and solved \eqref{eq9}, we can obtain
\begin{subequations}\label{eq34}
  \begin{align}
    {R}_{\phi,co}(\theta_{\mathrm{in}}) = \frac{{D}_{\phi,1}}{{D}_{\phi,0}}, \quad {R}_{\phi,cro}(\theta_{\mathrm{in}}) = \frac{{D}_{\phi,2}}{{D}_{\phi,0}},\\
    {T}_{\phi,co}(\theta_{\mathrm{in}}) = \frac{{D}_{\phi,3}}{{D}_{\phi,0}}, \quad  {T}_{\phi,cro}(\theta_{\mathrm{in}}) = \frac{{D}_{\phi,4}}{{D}_{\phi,0}},
  \end{align}
\end{subequations}
where ${D}_{\phi, i}$ is an abbreviation, and its specific expression can be found in the appendix A, ${R}_{\phi,co}(\theta_{\mathrm{in}})$ and ${T}_{\phi,co}$ represent the actual co-polarized reflection and transmission coefficients from $\mathbf{\hat{\phi}}$-polarized components to $\mathbf{\hat{\phi}}$-polarized components, ${R}_{\phi,cro}(\theta_{\mathrm{in}})$ and ${T}_{\phi,cro}$ represent the actual cross-polarized reflection and transmission coefficients from $\mathbf{\hat{\phi}}$-polarized components to $\mathbf{\hat{\theta}}$-polarized components, respectively.

As a result, the conversion from the predefined desired reflection/transmission coefficients to the actual reflection/transmission coefficients at a given incident angle for Type 3 RIS elements is achieved.
\subsection{RIS elements EM response}
Based on the actual reflection and transmission coefficients of RIS elements at a given incident angle, the EM field distribution on the RIS elements surfaces can be calculated. Furthermore, by combining the equivalence principle to calculate the secondary radiation of RIS elements, the EM response of RIS elements can be obtained. Takes the $G^{\theta \phi}$ in \eqref{eq3} as an example to illustrate the detailed process.

Assume that $\mathbf{\hat{\phi}}$-polarized signals $(\mathbf{E}_{i}, \mathbf{H}_{i})$ are incident at angles $(\theta_{\mathrm{in}}, \phi_{\mathrm{in}})$. With this information, the wave vectors of the reflected and transmitted signals can be calculated as
\begin{subequations}\label{eqn-18}
  \begin{align}
    &\mathbf{k}_{r} = \frac{2\pi}{\lambda}\begin{bmatrix}
-sin\theta_{\mathrm{in}}cos\phi_{\mathrm{in}}\\
-sin\theta_{\mathrm{in}}sin\phi_{\mathrm{in}}\\
cos\theta_{\mathrm{in}}
\end{bmatrix}
,\\
&\mathbf{k}_{t} = \frac{2\pi}{\lambda}\begin{bmatrix}
-sin\theta_{\mathrm{in}}cos\phi_{\mathrm{in}}\\
-sin\theta_{\mathrm{in}}sin\phi_{\mathrm{in}}\\
-cos\theta_{\mathrm{in}}
\end{bmatrix}.
  \end{align}
\end{subequations}
Then, the reflected fields and transmitted fields on the surface of the RIS element can be calculated.
\subsubsection{Type 1 RIS element}
The reflected fields on the RIS element surface are calculated as
\begin{subequations}\label{eq29}
  \begin{align}
    &\mathbf{E}_{r} = R_{\phi}(\theta_{\mathrm{in}})\mathbf{E}_{i},\label{eq28a}\\ &\mathbf{H}_{r} = \frac{\mathbf{k}_{r} \times \mathbf{E}_{r}}{|\mathbf{k}_{r}|\eta},\label{eq28b}
  \end{align}
\end{subequations}
and the transmitted fields on the RIS element surface are calculated as
\begin{equation}\label{eqn-18}
\mathbf{E}_{t} = T_{\phi}(\theta_{\mathrm{in}})\mathbf{E}_{i},\quad \mathbf{H}_{t} = \frac{\mathbf{k}_{t} \times \mathbf{E}_{t}}{|\mathbf{k}_{t}|\eta}.
\end{equation}
\subsubsection{Type 2 RIS element} The reflected electric field on the RIS element surface is calculated as
\begin{subequations}\label{eqn-18}
  \begin{align}
    &\mathbf{E}_{r} = 
    \begin{bmatrix}
R_{\phi,x}(\theta_{\mathrm{in}})E_{i,x}\\
R_{\phi,y}(\theta_{\mathrm{in}})E_{i,y}\\
E_{r,z}\\
\end{bmatrix},
  \end{align}
\end{subequations}
where $E_{i,x}$ and $E_{i,y}$ represent the $x$- and $y$-polarized components of $\mathbf{E}_{i}$, respectively. The $z$-polarized component of $\mathbf{E}_{r}$ is determined by $\mathbf{k}_{r}$ through the following relationship:
\begin{subequations}\label{eqn-18}
  \begin{align}
    &\mathbf{k}_{r}^{T}\cdot \mathbf{E}_{r} = 0.
  \end{align}
\end{subequations}
The reflected magnetic field is calculated using \eqref{eq28b}. 

Similarly, the transmitted EM field $(\mathbf{E}_{t}, \mathbf{H}_{t})$ can be calculated based on $T_{\phi,x}(\theta_{\mathrm{in}})$ and $T_{\phi,y}(\theta_{\mathrm{in}})$.
\subsubsection{Type 3 RIS element}
The reflected electric field on the Type 3 RIS element surface is calculated as
\begin{subequations}\label{eqn-18}
  \begin{align}
    &\mathbf{E}_{r} = ({R}_{\phi,co}(\theta_{\mathrm{in}})\mathbf{\hat{e}}^{\phi}_{r}\mathbf{\hat{e}}^{\phi}_{i}+{R}_{\phi,cro}(\theta_{\mathrm{in}})\mathbf{\hat{e}}^{\theta}_{r}\mathbf{\hat{e}}^{\phi}_{i})\mathbf{E}_{i},
  \end{align}
\end{subequations}
where $\mathbf{\hat{e}}^{\phi}_{r}\mathbf{\hat{e}}^{\phi}_{i}$ and $\mathbf{\hat{e}}^{\theta}_{r}\mathbf{\hat{e}}^{\phi}_{i}$ are dyadics, $\mathbf{\hat{e}}^{\phi}_{r}$ and $\mathbf{\hat{e}}^{\theta}_{r}$ represent the direction vector of $\mathbf{\hat{\phi}}$ and $\mathbf{\hat{\theta}}$ polarization for the reflected signal, and $\mathbf{\hat{e}}^{\phi}_{i}$ represents the direction vector of $\mathbf{\hat{\phi}}$ polarization for the incident signal. The reflected magnetic field is calculated using \eqref{eq28b}. 

Similarly, the transmitted EM fields $(\mathbf{E}_{t}, \mathbf{H}_{t})$ can be calculated based on ${T}_{\phi,co}(\theta_{\mathrm{in}})$ and ${T}_{\phi,cro}(\theta_{\mathrm{in}})$.

We have now determined the reflected and transmitted fields on the RIS element surface. More importantly, this is based on the actual reflection/transmission coefficients, rather than directly employing the predefined desired ones ($ \overline{\overline{\mathbf{R}}}$ and $ \overline{\overline{\mathbf{T}}}$). This is because when the signal is obliquely incident, $ \overline{\overline{\mathbf{R}}}$ and $ \overline{\overline{\mathbf{T}}}$ may become invalid. According to the equivalence principle (akin to Huygens' principle), the RIS element surface with its EM field distribution can be considered as a secondary radiation source. 
Specifically, in the reflective half-space, the equivalent radiation sources can be calculated as
\begin{subequations}\label{eqn-18}
  \begin{align}
    \mathbf{J}_{r} = \mathbf{\hat{z}}\times (\mathbf{H}_{i}+\mathbf{H}_{r}),\\
    \mathbf{M}_{r} =-\mathbf{\hat{z}} \times (\mathbf{E}_{i}+\mathbf{E}_{r}),
  \end{align}
\end{subequations}
where $\mathbf{J}_{r}$ and $\mathbf{M}_{r}$ are equivalent radiation EM current density, respectively.
And in the transmissive half-space, the equivalent radiation source is calculated as
\begin{equation}\label{fundamental_diode_voltage}
    \mathbf{J}_{t} = \mathbf{\hat{z}}\times \mathbf{H}_{t},  \quad \mathbf{M}_{t} =- \mathbf{\hat{z}} \times \mathbf{E}_{t}.\\
\end{equation}

Thus, $G^{\theta \phi}$ can be obtained by calculating the radiation of $\mathbf{J}_{r/t}$ and $\mathbf{M}_{r/t}$. To this end, we choose an observation point $O$  located in the far-field of the RIS element in the target direction $(\theta_{\mathrm{out}}, \phi_{\mathrm{out}})$ and record the distance from $O$ to the RIS element center as $r$. Then, the specific process is as follows:

Firstly, the potential vector $\mathbf{A}$ and $\mathbf{F}$ are calculated as
\begin{subequations}\label{eqn-18}
  \begin{align}
    \mathbf{A} =\frac{\mu_{0}}{4\pi}\iint_{S} \mathbf{J}_{r/t}\frac{e^{-jkR}}{R}d s^{'},\\
    \mathbf{F} =\frac{\epsilon_{0}}{4\pi}\iint_{S} \mathbf{M}_{r/t}\frac{e^{-jkR}}{R}d s^{'},
  \end{align}
\end{subequations}
where $\mu_{0}$ and $\epsilon_{0}$ represent the permeability and  permittivity in vacuum, $S$ represents the RIS element surface, and $R$ denotes the distance from  $O$ to $d s^{'}$ \cite{2013Advanced}.
Note that if the observation point $O$ is located in the reflective half-space, the equivalent radiation EM current density should be  $\mathbf{J}_{r}$ and $\mathbf{M}_{r}$. Otherwise, $\mathbf{J}_{t}$ and $\mathbf{M}_{t}$ should be utilized. 

Then,  the radiation electric field $\mathbf{E}_{s}$ at $O$ generated by the equivalent EM currents can be calculated as
\begin{subequations}\label{eqn-18}
  \begin{align}
\mathbf{E}_{s}\approx-j\omega \mathbf{A}-\frac{1}{\epsilon_{0}}\nabla \times \mathbf{F},
  \end{align}
\end{subequations}
where $\omega = 2\pi f$ and $f$ is the signal frequency.

Finally, the RIS element EM response can be calculated as
\begin{equation}\label{eqn-18}
G^{\theta\phi} = \lim_{r\rightarrow \infty}\frac{4\pi}{\lambda}re^{j2\pi \lambda^{-1}r}\frac{E_{s,\theta}}{E_{i}},
\end{equation}
where $E_{s,\theta}$ represent the $\mathbf{\hat{\theta}}$-polarized components of $\mathbf{E}_{s}$, $E_{i}$ is the scalar form of $\mathbf{E}_{i}$. According to the previous assumption, $\mathbf{E}_{i}$ is purely $\mathbf{\hat{\phi}}$-polarized. The coefficient $\frac{4\pi}{\lambda}$ results from converting electric field amplitude to multipath amplitude \cite{ghw}. $re^{j2\pi \lambda^{-1}r}$ compensates for the attenuation caused by the distance between the RIS element and $O$. This ensures that the final radiation pattern is independent of $r$.

\subsection{Passive and lossless conditions}
To reduce the hardware cost of RIS elements and improve the energy efficiency of the communication system, the RIS elements are expected to be designed as passive and lossless. Furthermore, the passive and lossless conditions also serve as the prerequisite for a fair performance comparison of different types of RIS. For impedance tensors, a passive and lossless RIS element should satisfy \eqref{eq10}. For the predefined desired reflection and transmission coefficients, it should fulfill
\begin{equation}\label{fundamental_diode_voltage}
\begin{bmatrix}
\overline{\overline{\mathbf{R}}} & \overline{\overline{\mathbf{T}}} \\
\overline{\overline{\mathbf{T}}} & \overline{\overline{\mathbf{R}}} \\
\end{bmatrix}^{T} \cdot
\begin{bmatrix}
\overline{\overline{\mathbf{R}}} & \overline{\overline{\mathbf{T}}} \\
\overline{\overline{\mathbf{T}}} & \overline{\overline{\mathbf{R}}} \\
\end{bmatrix}^{*} = \overline{\overline{\mathbf{I}}},
\end{equation}
where $\overline{\overline{\mathbf{I}}}  \in \mathbb{C} ^{4 \times 4}$  is identity matrix \cite{GTSC2}. Then, the passive and lossless conditions of each type of RIS element can be obtained respectively.

\subsubsection{Type 1 RIS element}
If a type 1 RIS element is passive and lossless, its predefined reflection and transmission coefficients should meet 
\begin{subequations}\label{eqn-18}
  \begin{align}
 |R|^{2}+|T|^{2} = 1,\\
\angle R-\angle T = \frac{\pi}{2}+v\pi,
  \end{align}
\end{subequations}
where $v \in \{0, 1\}$ \cite{STAR-RIS}. 
\subsubsection{Type 2 RIS element} 
If a type 2 RIS element is passive and lossless, its predefined desired reflection and transmission coefficients should meet the following conditions:
\begin{subequations}\label{eqn-18}
  \begin{align}
  |R_{a}|^{2}+|T_{a}|^{2} = 1,\\
 \angle R_{a}-\angle T_{a} = \frac{\pi}{2}+v_{a}\pi,
  \end{align}
\end{subequations}
where $a \in \{x,y\}$, $v_{a} \in\{0,1\}$.
\subsubsection{Type 3 RIS element}
If a type 3 RIS element is passive and lossless, its predefined desired reflection and transmission coefficients should meet the following conditions:
\begin{subequations}\label{eqn-18}
  \begin{align}
&|R_{co}|^{2}+|R_{cro}|^{2}+|T_{co}|^{2}+|T_{cro}|^{2} = 1,\\
&\angle\Gamma_{c}-\angle\Psi_{d}=
\begin{cases}  
\frac{\pi}{2}+v_{cd}^{\Gamma\Phi}\pi, & \text{if} \quad \Gamma \neq \Psi\quad \text{and} \quad c = d\\  
\pi+v_{cd}^{\Gamma\Phi}\pi,  & \text{otherwise}
\end{cases},
  \end{align}
\end{subequations}
where $\Gamma, \Psi \in \{R,T\}$.  And $c, d \in \{co, cro\}$, $v_{cd}^{\Gamma\Phi} \in \{0,1\}$.

\section{numerical analysis and validation}
In this section, we first investigate the incident angle dependence of RIS element reflection and transmission coefficients and its impact on the RIS performance in the free space scenario. Subsequently, the performance of different types of RIS is compared in a T-shaped corridor under passive and lossless conditions. 

\begin{figure}[!htb]
    \centering
    \subfigure[]{\includegraphics[width=0.45\hsize, height=0.3\hsize]{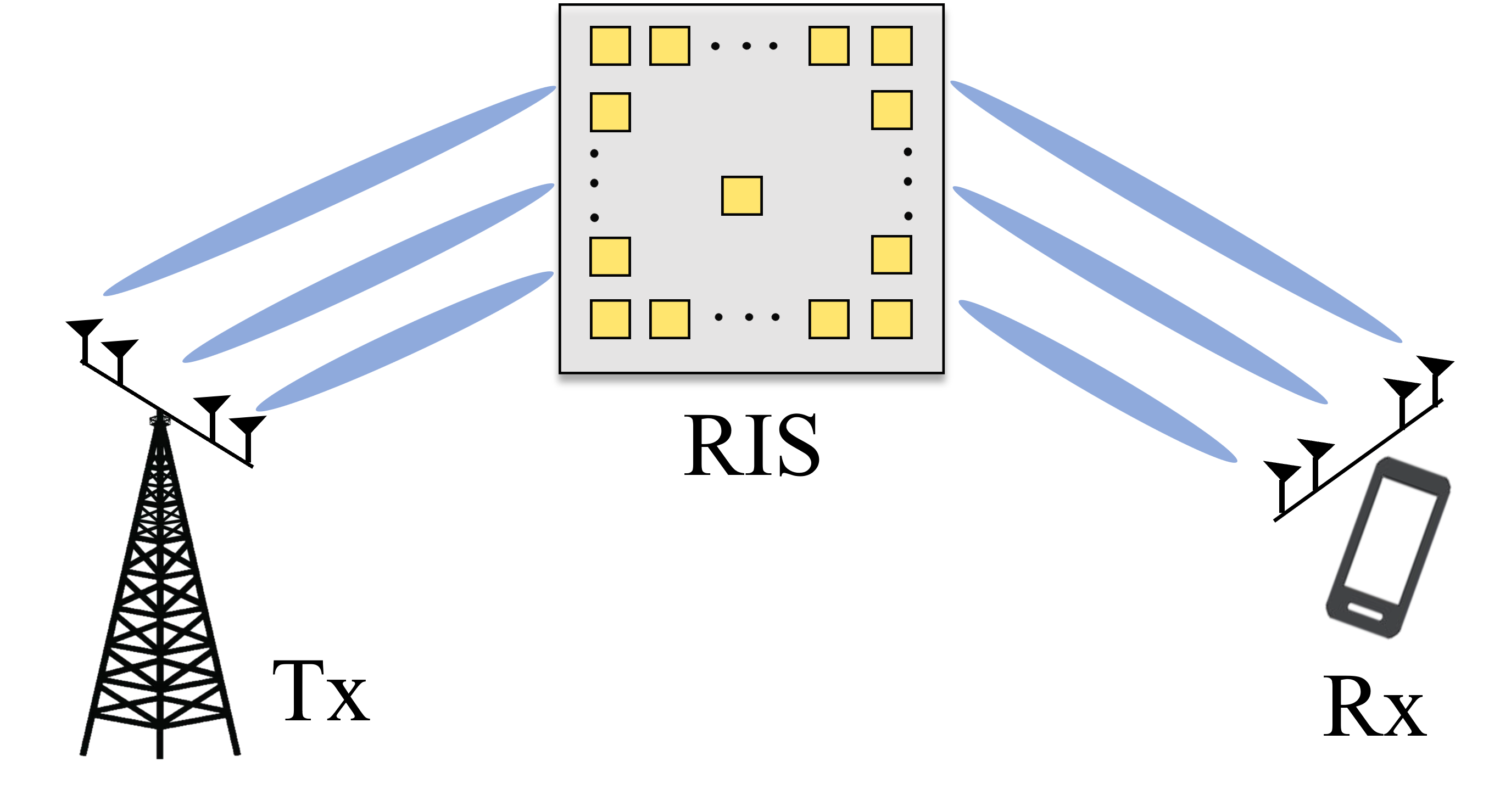}\label{fig: sub_figure1}}
    \subfigure[]{\includegraphics[width=0.45\hsize, height=0.3\hsize]{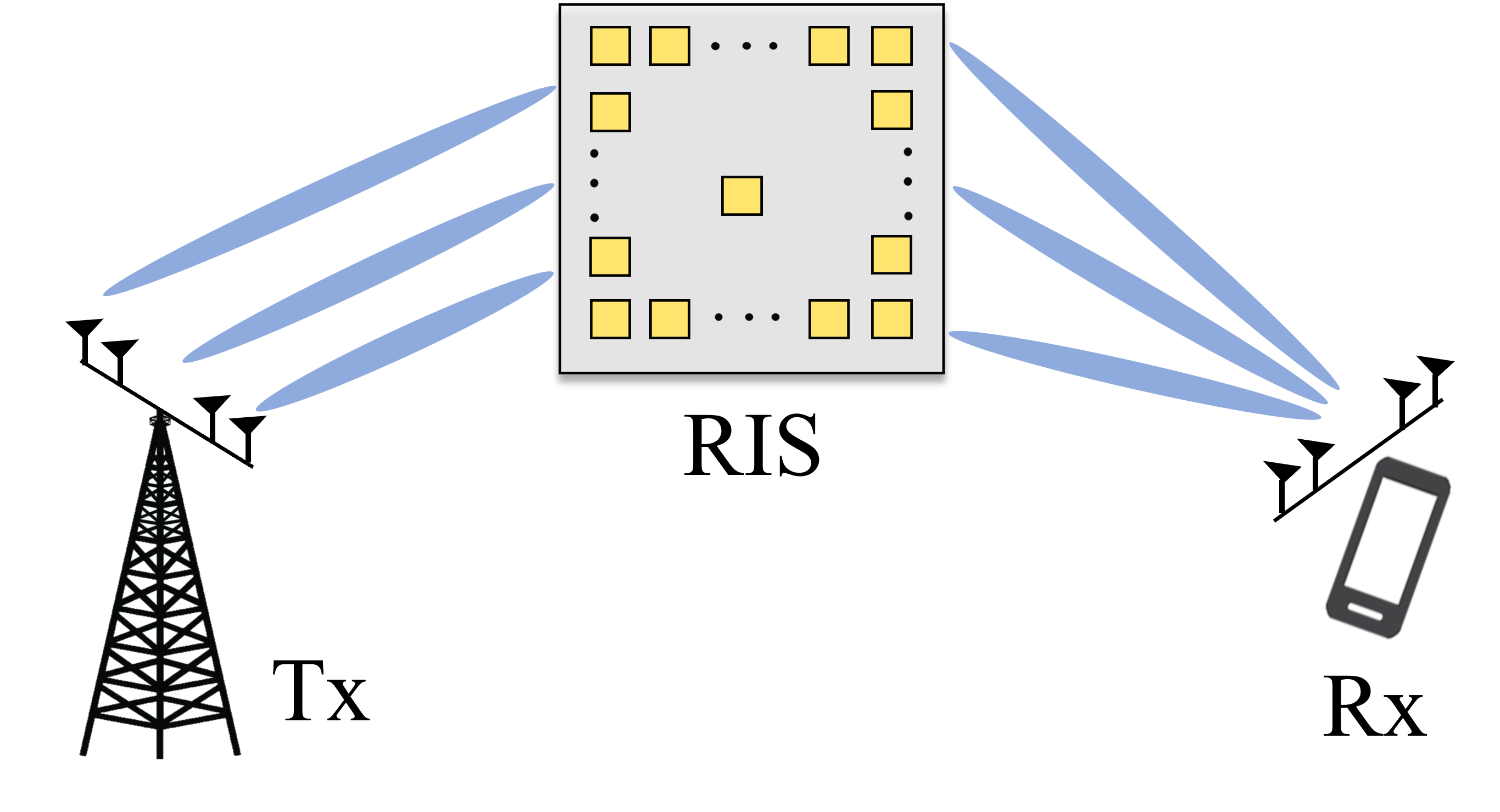}\label{fig: sub_figure2}}
    \caption{Several functions of RIS: a) beamforming. b) beamfocusing.}
    \label{RIS_function}
\end{figure}
In all simulations, the following two codebooks and their variants are mainly used \cite{smartRadio}:

Codebook1: This codebook enables the RIS to perform the beamforming function, steering the reflected/transmitted beam towards the target direction, as depicted in Fig. \ref{RIS_function}(a). It is suitable for the case where both Tx and Rx are located in the RIS far-field.

Codebook2: This codebook enables the RIS to perform the beamfocusing function, concentrating the reflected/transmitted power to a single point, as illustrated in Fig. \ref{RIS_function}(b). It is suitable for the case where Tx is located in the RIS far-field while Rx is located in the RIS near-field.
\subsection{Deviation of actual reflection/transmission coefficients}
To utilize RIS for various functions, such as enhancing coverage and improving channel rank conditions, each RIS element must be configured with an appropriate reflection/transmission coefficient. However, as a large number of RIS elements are designed for signals incident at normal angles, their actual reflection/transmission coefficients can deviate from predefined desired values when the signals are obliquely incident (see subsection B of Section III for details). Next, we examine the impact of this deviation on the RIS elements and panels.

First, the phase deviation of the actual reflection/transmission coefficients of RIS elements is studied, which is a crucial manifestation of the incident angle dependence.  
Taking a dual-polarized unified control RIS element as an example, suppose its operating center frequency is 26 GHz and its size is $\frac{\lambda}{2} \times \frac{\lambda}{2}$. Assume its predefined desired reflection and transmission coefficient are $R=|R|e^{j\angle R}$ and  $T=|T|e^{j\angle T}$. To satisfy the passive and lossless conditions, let $|R|=|T|=\frac{\sqrt{2}}{2}$ and $\angle T = \angle R + \frac{\pi}{2}$. Thus, when $\mathbf{\hat{\phi}}$-polarized signals are incident at an angle $(\theta_{\mathrm{in}},\phi_{\mathrm{in}})$, the phase deviation of the actual reflection/transmission coefficients can be represented as
\begin{subequations}\label{eqn-18}
  \begin{align}
\varphi_{\phi}^{R}(\theta_{\mathrm{in}}) = |\angle R_{\phi}(\theta_{\mathrm{in}}) - \angle R| 
,\\
\varphi_{\phi}^{T}(\theta_{\mathrm{in}}) = |\angle T_{\phi}(\theta_{\mathrm{in}}) - \angle T|.
  \end{align}
\end{subequations}
Here,  $R_{\phi}(\theta_{\mathrm{in}})$ and $T_{\phi}(\theta_{\mathrm{in}})$ are calculate according to \eqref{eq28} in subsection B of Section III. 

\begin{figure}[!htb]
    \centering
    \subfigure[]{\includegraphics[width=0.49\hsize, height=0.5\hsize]{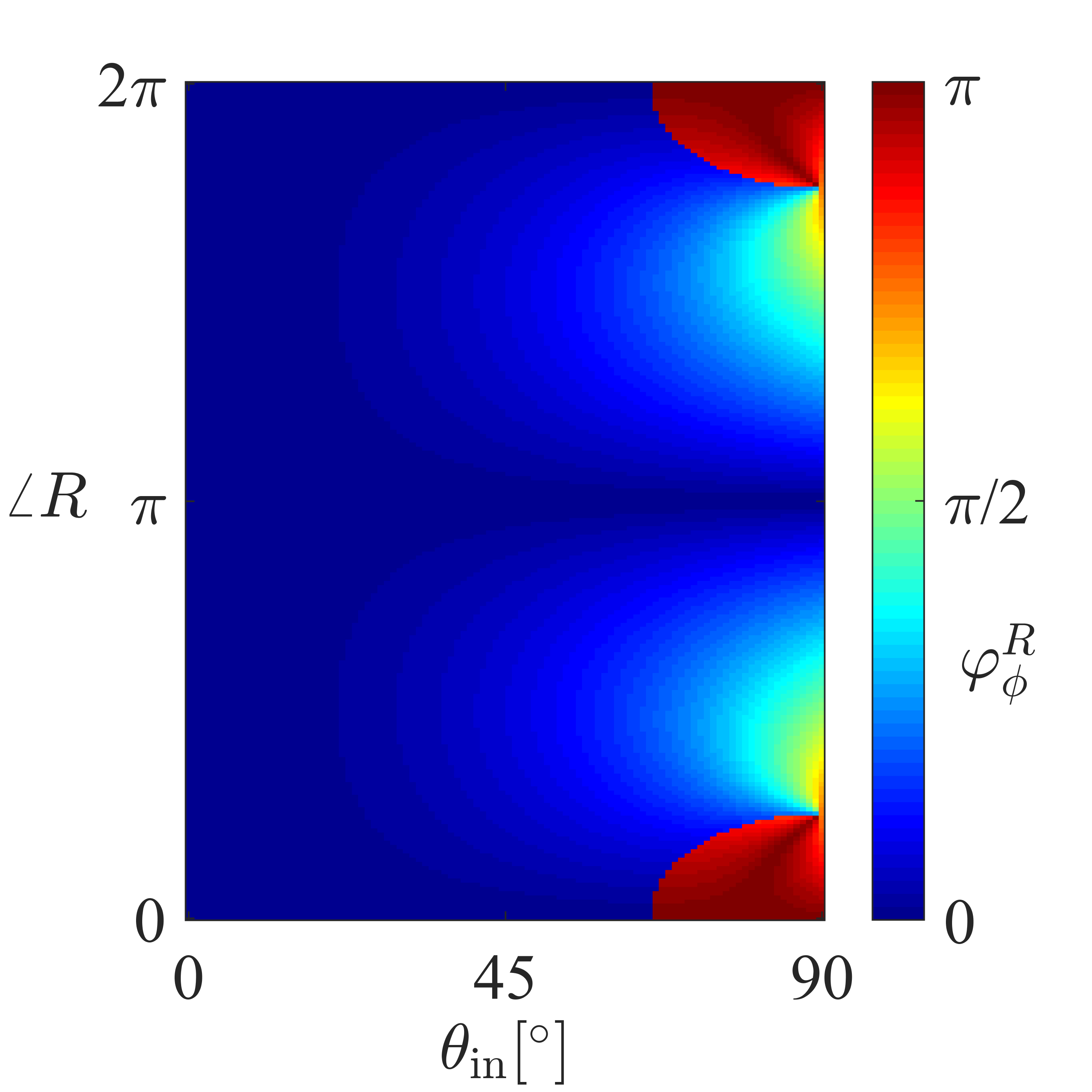}\label{fig: sub_figure1}}
    \subfigure[]{\includegraphics[width=0.49\hsize, height=0.5\hsize]{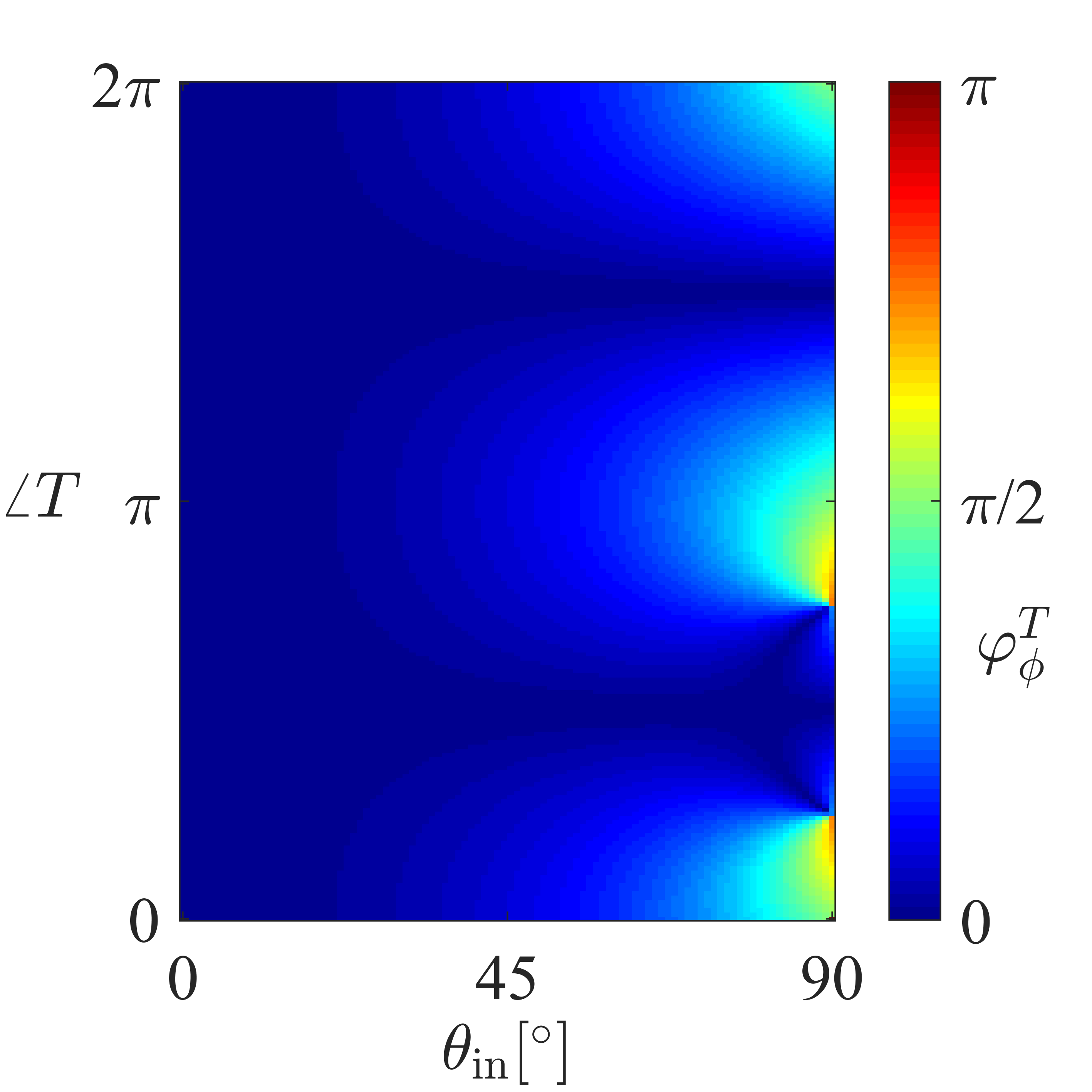}\label{fig: sub_figure2}}
    \caption{The phase deviation of the actual reflection/transmission coefficient: a) The phase deviation of the actual reflection coefficient b) The phase deviation of the actual transmission coefficient}
    \label{ref_deviation}
\end{figure}

Fig. \ref{ref_deviation} shows the relationship between phase deviation, incident angle, and predefined desired reflection/transmission coefficient. It is evident that as the incident angle increases, so does the phase deviation from the predefined desired value. Specifically, when the incident angle $\theta_{\mathrm{in}}$ is less than $40^\circ$, the phase deviation is negligible, but when $\theta_{\mathrm{in}}$ exceeds $40^\circ$, the phase deviation becomes significant. Additionally, at the same incident angle, the phase deviation varies with the predefined desired reflection/transmission coefficient.

\begin{figure}
  \begin{center}
  \includegraphics[width=3.4in]{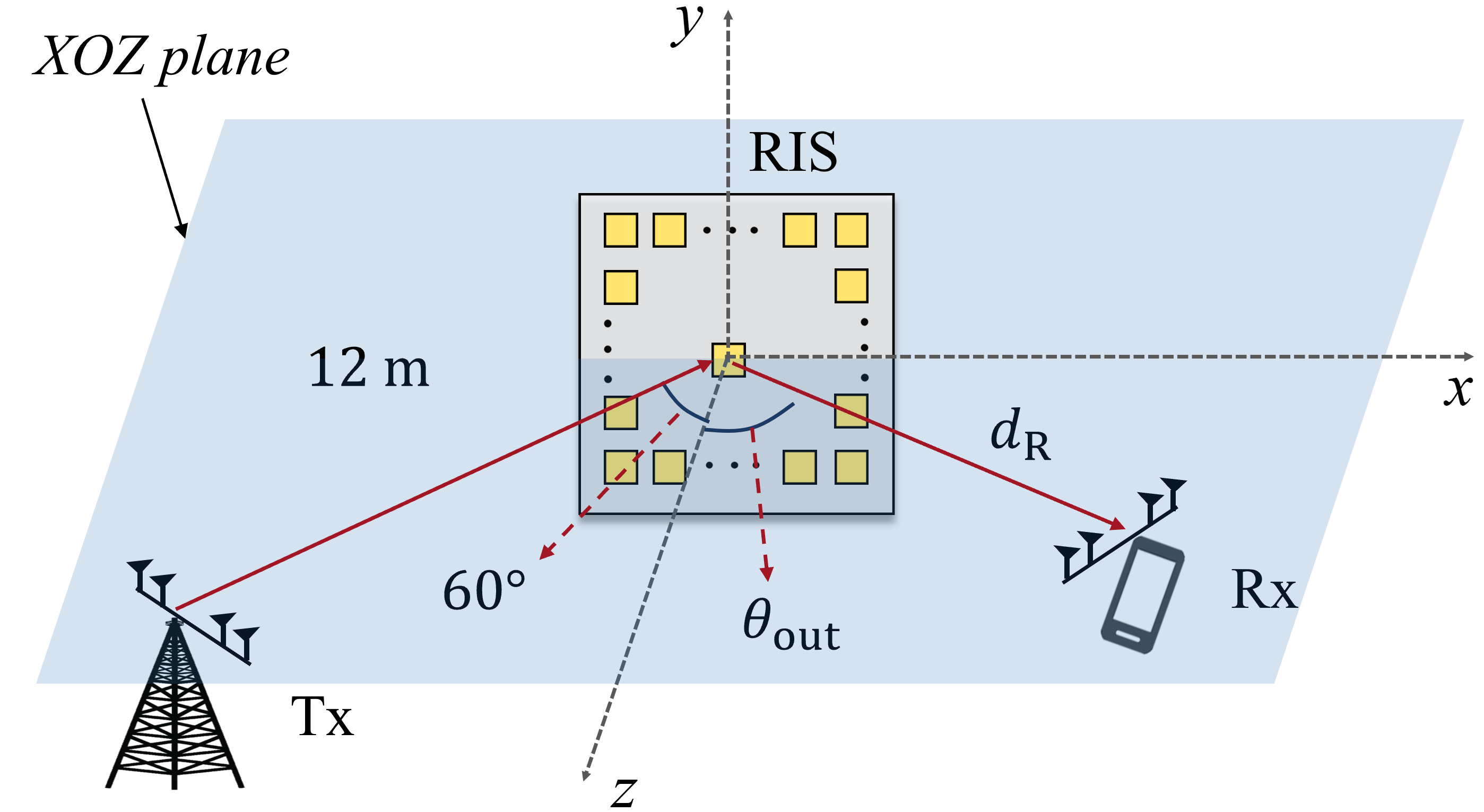}
  \caption{The free space simulation scenario}\label{free_space}
  \end{center}
\end{figure}

Then, the impact of the deviation of the actual reflection/transmission coefficient on the RIS performance is studied. Fig. \ref{free_space} shows the simulation scenario. In the scenario, Tx and Rx are placed on the $XOZ$ plane, with Rx located on the right side of the $YOZ$ plane and Tx located on the left side of the $YOZ$ plane. The zenith angle of Tx relative to the RIS is $60^\circ$, and the distance between Tx and the RIS center is 12 m. The zenith angles of Rx relative to the RIS is $\theta_{\mathrm{out}}$, and the distance between Rx and RIS center is $d_{R}$. The RIS consisted of $30 \times 30$ elements and its operating frequency is 26 GHz.

Simulations are conducted in both near-field and far-field cases. The specific simulation configuration is as follows: Assuming that Tx and Rx are equipped with dual-polarized antennas, which are capable of transmitting and receiving dual-polarized signals. For the near-field case, $d_{R}$ is set to 2 m, and the RIS adopts the codebook2  to concentrate power at the point with $d_{R} =2$ m and $\theta_{\mathrm{in}} = 45^\circ$. For the far-field case, $d_{R}$ is set to 20 m, and the RIS adopts the codebook1 to direct the beam towards the direction of $\theta_{\mathrm{in}}=45^\circ$. 

\begin{figure}
  \begin{center}
  \includegraphics[width=3.4in]{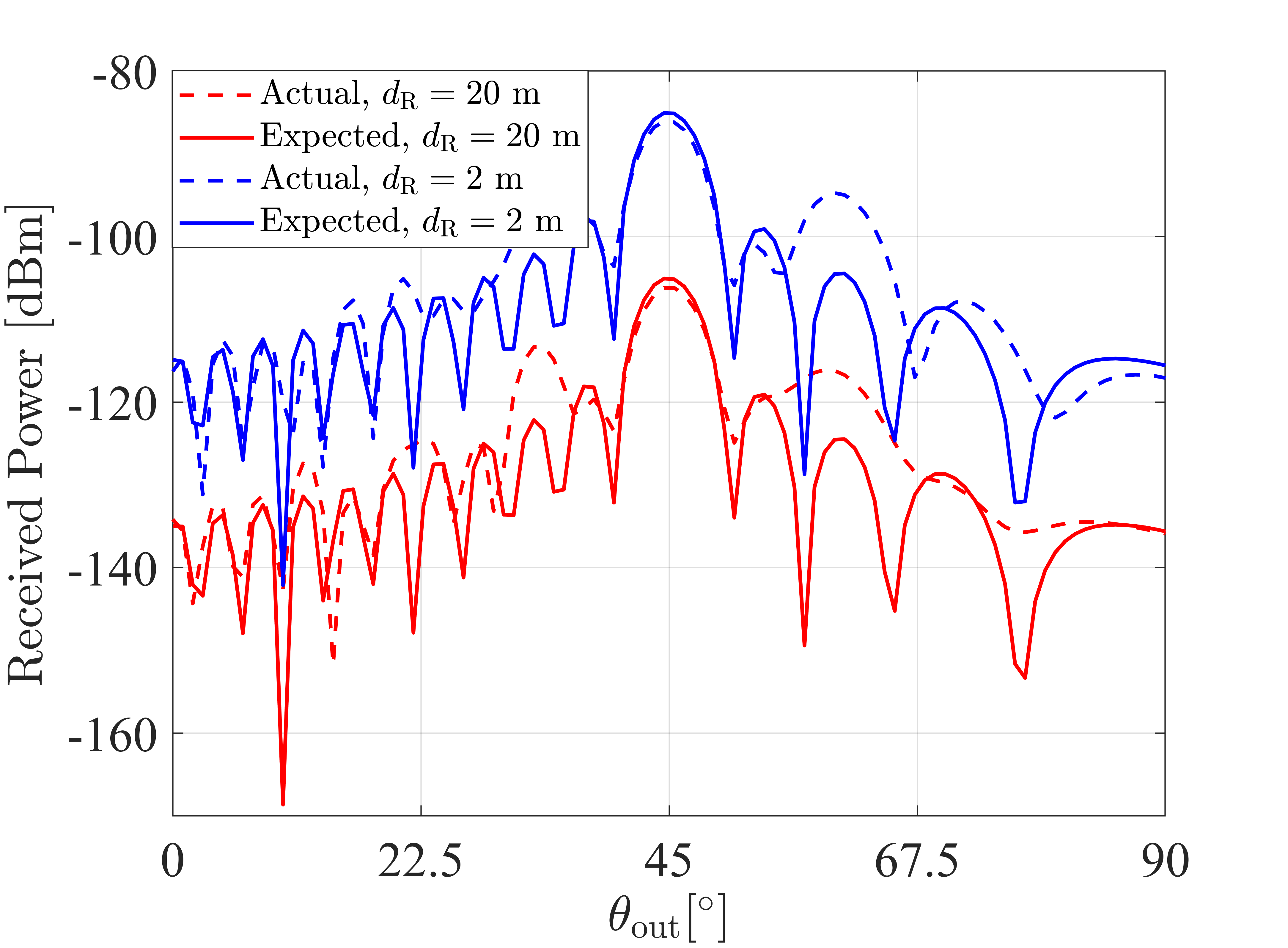}
  \caption{The impact of the actual reflection/transmission coefficient and the predefined desired reflection/transmission coefficient on the performance of RIS.}\label{power_deviation}
  \end{center}
\end{figure}

Fig. \ref{power_deviation} illustrates the received power of the Rx. The blue lines represent the near-field case, while the red lines represent the far-field case. 
The dashed lines represent the results that consider the incident angle dependence, where the EM response is calculated based on the actual reflection coefficients of the RIS elements, and the received power is further computed. The solid lines represent the desired results, where the EM response is calculated based on the predefined desired reflection coefficients, and the received power is further computed.
Firstly, it can be seen that when $\theta_{\mathrm{out}} = 45^\circ$, the Rx receives substantial power due to the critical role of beamforming and beamfocusing. Moreover, whether the Rx is in the far-field (red lines) or near-field (blue lines) of the RIS, the signal power calculated using the actual reflection coefficient (dashed lines) in the target beamforming direction or beamfocusing point is lower than that calculated with the predefined desired reflection coefficient (solid lines).  Conversely, in non-target beamforming directions or focus points, using the actual reflection coefficients results in higher sidelobes (dashed lines). This indicates that simulations using the predefined desired reflection coefficient may lead to an overestimation of RIS performance. In practice, the RIS might not achieve the desired effect and could inadvertently interfere with other users due to the increased sidelobe power.

\subsection{Performance comparison of different types of RIS}
\begin{figure}
  \begin{center}
  \includegraphics[width=3.4in]{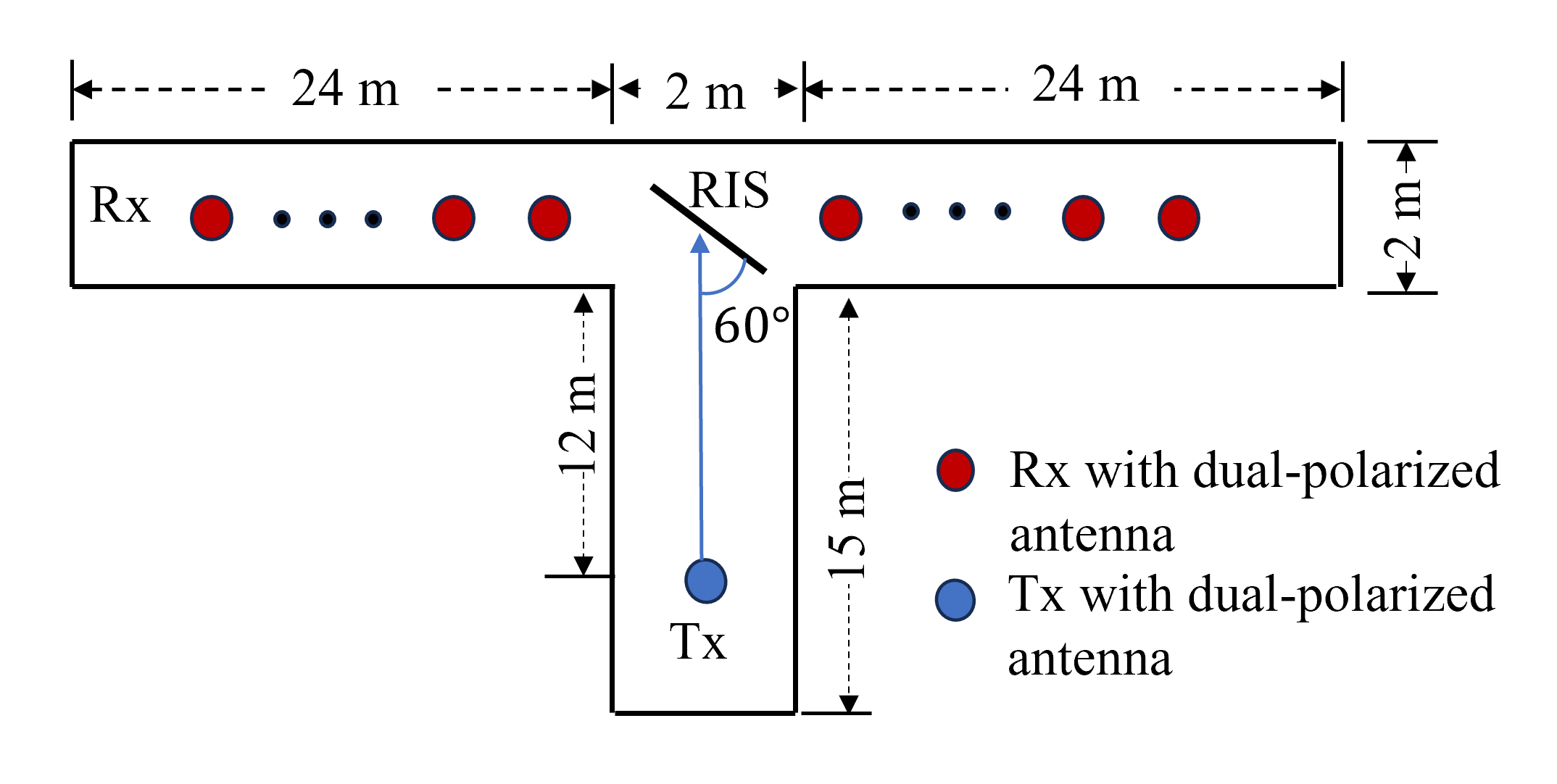}
  \caption{T-shaped corridor: the simulation scenario for performance comparison of dual-polarized unified control and dual-polarized independent control RIS}\label{T_corridor1}
  \end{center}
\end{figure}

The following compares the performance of various types of RIS. Firstly, the performance of dual-polarized unified control and dual-polarized independent control RIS is compared. Fig. \ref{T_corridor1} shows the simulation scenario. It is a T-shaped corridor, composed of a horizontal corridor and a vertical corridor. The RIS is deployed at the corner of the corridor, forming a $60^\circ$ angle with the vertical corridor. The Tx is deployed on the vertical corridor, 12 m away from the RIS, and a large number of Rx are deployed on the horizontal corridor, distributed on both the reflecting and transmitting sides of the RIS. All of them are positioned at a height of 1.5 m. 

The Tx is equipped with a dual-polarized antenna and the power proportion of $\mathbf{\hat{\theta}}'$- and $\mathbf{\hat{\phi}}'$-polarized components is adjusted such that the LOS path in the Tx-RIS subchannel has identical $\mathbf{\hat{\theta}}$- and $\mathbf{\hat{\phi}}$-polarized components. The Rx is also equipped with dual-polarized antennas to receive dual-polarized signals. The RIS consists of $M \times M$ elements and operates at a frequency of 26 GHz.

For dual-polarized unified control RIS, we first set $|R| = |T|=\frac{\sqrt{2}}{2}$ for each element to enable the RIS to serve both the reflecting and the transmitting side simultaneously \cite{STAR-RIS}\cite{lixinagwang}. Then, $\angle R$ and $\angle T$ of each element are configured according to codebook1, directing the reflected beam toward Rx on the reflecting side and the transmitted beam toward Rx on the transmitting side. Finally,  $\angle R$ and $\angle T$ of each element are adjusted to satisfy the passive and lossless conditions. The method of adjustment is detailed in Table \ref{tab1} \cite{STAR-RIS}, where $\phi_{m}$ is the mean value of $\angle R$ and $\angle T$ for $m$-th element, and $\angle R'$ and $\angle T'$ are the adjusted results.
\begin{table}[h]
\caption{Phase adjustment \cite{STAR-RIS}}
\centering
\begin{tabular}{|c|c|c|c|c|}
\hline
$\angle R - \angle T$ & $[0, \pi)$ & $[\pi, 2\pi)$ & $[-\pi, 0)$ & $[-2\pi, -\pi)$\\
\hline
$\angle R'$ & $\phi_{m}+\frac{\pi}{4}$ & $\phi_{m}+\frac{3\pi}{4}$
 & $\phi_{m}-\frac{\pi}{4}$& $\phi_{m}-\frac{3\pi}{4}$\\
\hline
$\angle T'$ & $\phi_{m}-\frac{\pi}{4}$ & $\phi_{m}-\frac{3\pi}{4}$
 & $\phi_{m}+\frac{\pi}{4}$ & $\phi_{m}+\frac{3\pi}{4}$\\
 \hline
\end{tabular}
\label{tab1}
\end{table}

For dual-polarized independent control RIS, it adopts the codebook1 to fully reflect the $x$-polarized signal component towards Rx on the reflecting side and fully transmit the $y$-polarized signal component towards Rx on the transmitting side. 

\begin{figure}
  \begin{center}
  \includegraphics[width=3.4in]{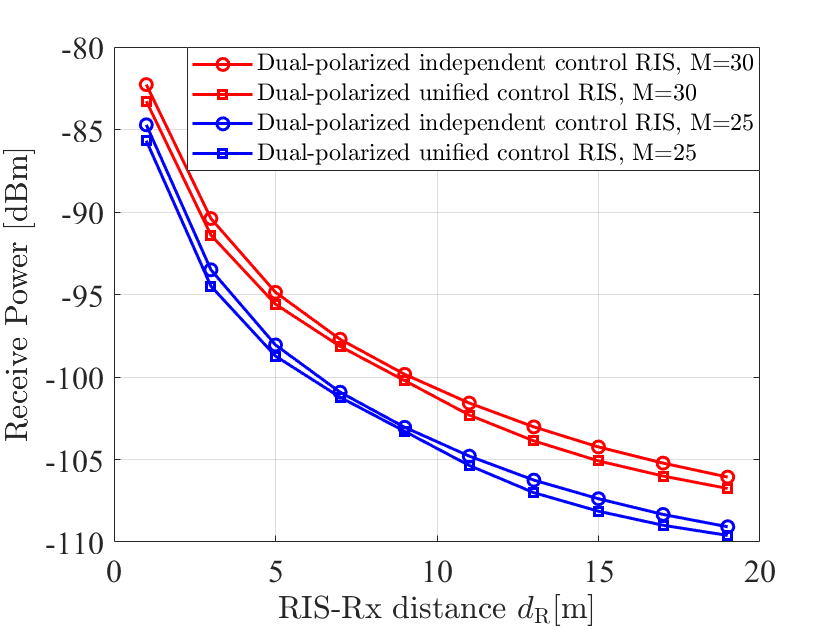}
  \caption{Performance comparison of dual-polarized unified control RIS and dual-polarized independent control RIS at RIS reflecting side}\label{single_vs_dual}
  \end{center}
\end{figure}

\begin{figure}
  \begin{center}
  \includegraphics[width=3.4in]{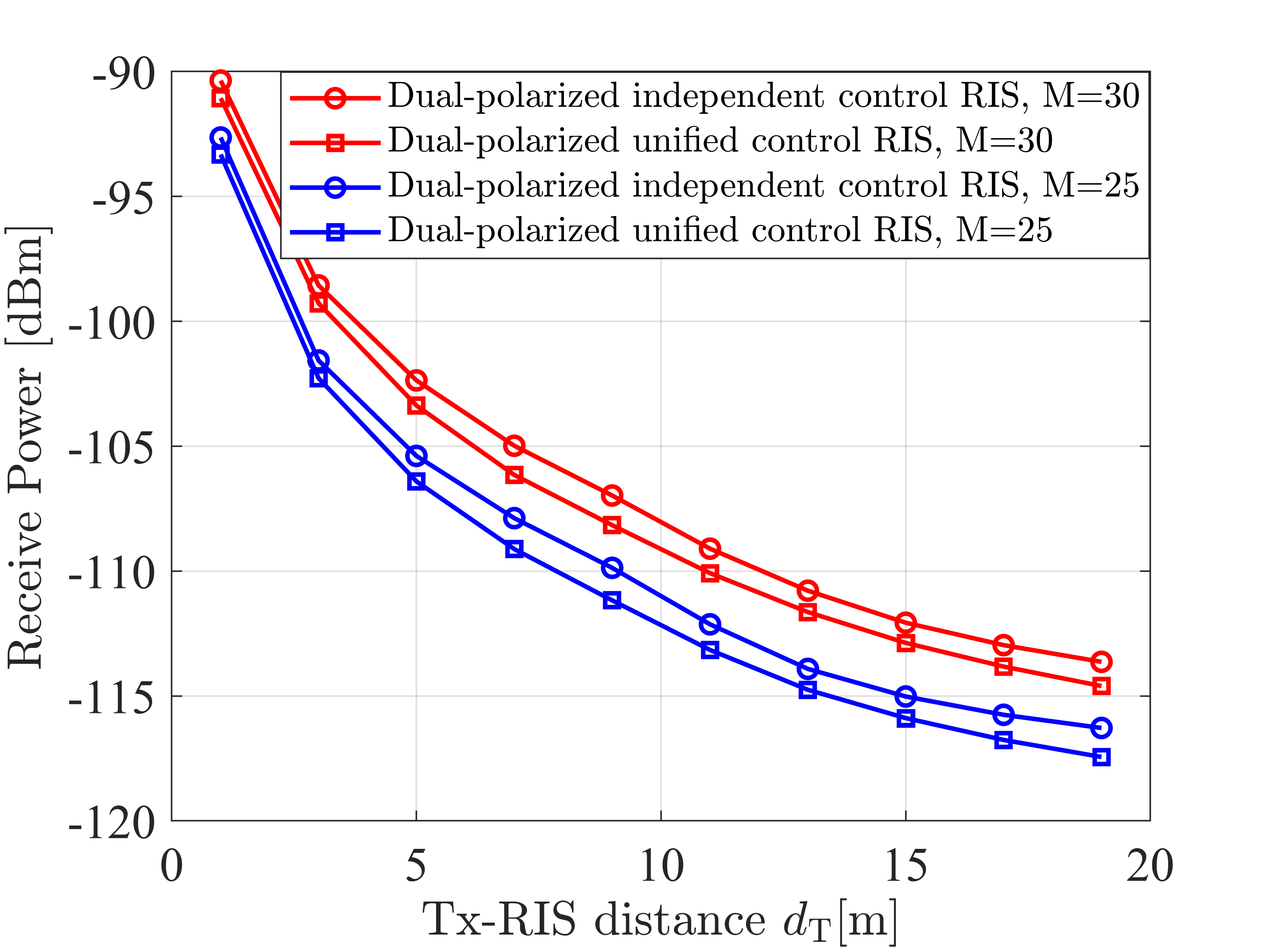}
  \caption{Performance comparison of dual-polarized unified control RIS and dual-polarized independent control RIS at RIS transmitting side}\label{single_vs_dual1}
  \end{center}
\end{figure}

Set the maximum reflection number of RT to 1 and the simulation frequency to 26 GHz. Fig. \ref{single_vs_dual} and Fig. \ref{single_vs_dual1} illustrate the received power of Rx in the reflecting and transmitting sides, respectively. It is apparent that the dual-polarized independent control RIS outperforms the dual-polarized unified control RIS, regardless of the reflection side or the transmission side. This is because the dual-polarized unified control RIS is constrained by passive and lossless conditions so that the phase of its reflection and transmission coefficients can not be freely configured by the user. Conversely, the dual-polarized independent control RIS ingeniously circumvents these conditions by independently controlling the  $x$-polarized and $y$-polarized components, thus allowing free configuration of the phases of its reflection and transmission coefficients.

\begin{figure}
  \begin{center}
  \includegraphics[width=3.3in]{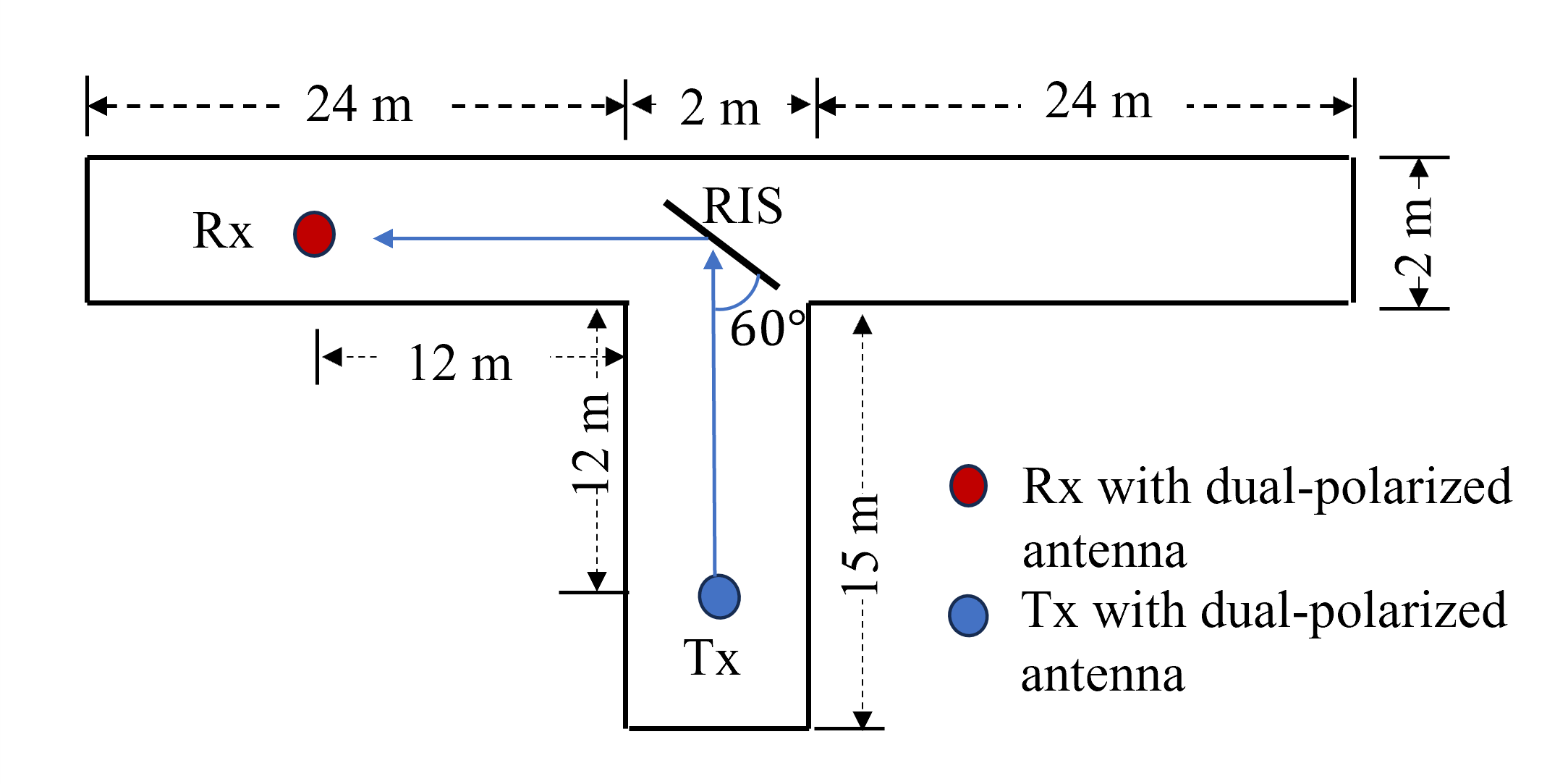}
  \caption{T-shaped corridor: the simulation scenario for performance comparison of dual-polarized unified control and polarization-rotation RIS}\label{T_corridor2}
  \end{center}
\end{figure}

Then, the performance of dual-polarized unified control RIS and polarization-rotation RIS is compared. The simulation scenario is shown in Fig \ref{T_corridor2}. The RIS is deployed at the corner of the corridor and consists of $ 30\times 30$ elements. The Tx is deployed on the vertical corridor, 12 m away from the RIS. It is equipped with a dual-polarized antenna, where the power proportion of $\mathbf{\hat{\theta}}'$-polarized signals is $\alpha$. The Rx is deployed on the horizontal corridor, located on the reflecting side of the RIS and 12 m away from it. It is equipped with a vertically polarized antenna that can only receive $\mathbf{\hat{\theta}}'$-polarized signals.

The dual-polarized unified control RIS adopts codebook1 to direct the reflected beam to the Rx. The polarization-rotating RIS works in three cases successively. The first is the case without polarization rotation, where $|R_{co}|$ of each element is set to 1 and $|R_{cro}|$ of each element is set to 0. $\angle R_{co}$ is configured according to codebook1, steering the reflected beam to the Rx. The second is the partial polarization rotation case, which sets  $|R_{co}|=|R_{cro}|=\frac{\sqrt{2}}{2}$ for each element, resulting in half of the incident $\mathbf{\hat{\theta}}$/$\mathbf{\hat{\phi}}$-polarized components being converted into $\mathbf{\hat{\phi}}$/$\mathbf{\hat{\theta}}$-polarized components. $\angle R_{co}$ is configured according to codebook1 to steer the co-reflected beam towards the Rx. And $\angle R_{cro}$ is set to $\angle R_{co}+\pi$ for each element to satisfy the passive lossness conditions. The third is the complete polarization rotation case, where $|R_{cro}|$ of each element is set to 1 and $|R_{co}|$ of each element is set to 0. $\angle R_{cro}$ is configured according to codebook1, steering the reflected beam to the Rx.

\begin{figure}
  \begin{center}
  \includegraphics[width=3.4in]{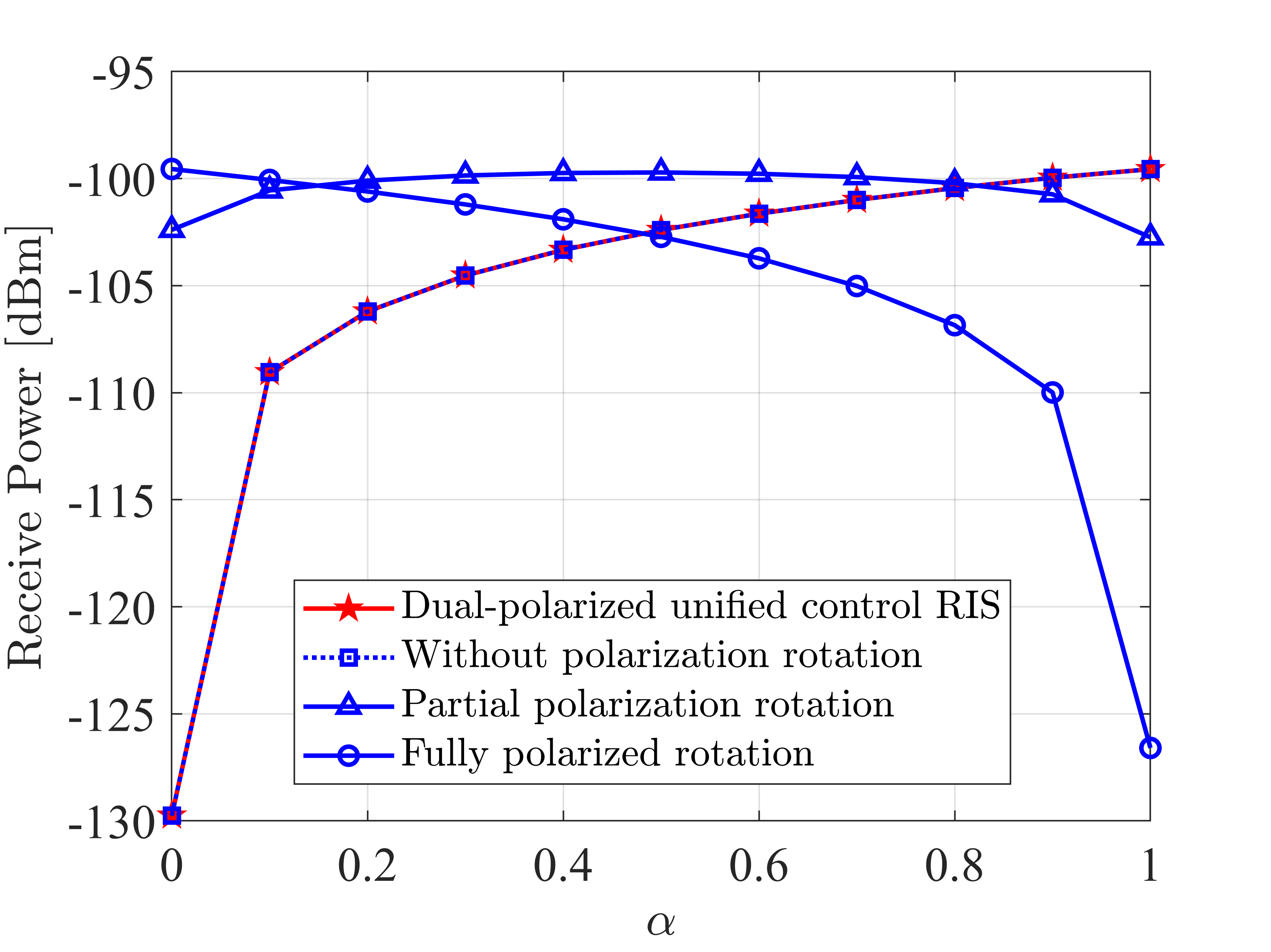}
  \caption{Performance comparison between dual-polarized unified control RIS and polarization-rotation RIS}\label{single_vs_rotation}
  \end{center}
\end{figure}
Set the maximum reflection number of RT to 1 and the simulation frequency to 26 GHz. Fig. \ref{single_vs_rotation} illustrates the relationship between the received power of the Rx and $\alpha$. It is evident that, due to the lack of polarization rotation capability in dual-polarized unified control RIS, the received power of the Rx increases with the increase of $\alpha$. When $\alpha$ is minimal, the Rx receives only a minimal amount of power. The performance of the polarization-rotating RIS aligns with that of the dual-polarized unified control RIS in the case without polarization rotation. In the complete polarization rotation case, the received power of the Rx decreases as $\alpha$ increases, showing a roughly symmetrical relationship to the curve of the dual-polarized unified control RIS.  
In the case of partial polarization rotation, the received power of the Rx can be consistently maintained at a high level, regardless of the value of $\alpha$. These results suggest that the codebook for polarization-rotating RIS can be flexibly selected according to the proportion of vertical and parallel polarization components, thus enhancing the received power for the user.

\section{Conclusion}
This paper establishes a unified and deterministic channel model for RIS in RT frameworks. The key contribution lies in the development of a generic RIS electromagnetic (EM) response model that seamlessly integrates into RT-based simulations, enabling the high-precision modeling of multiple RIS types, including dual-polarized unified control, dual-polarized independent control, and polarization-rotating RIS.

The proposed framework achieves this through three core steps. First, impedance-based GSTCs and the equivalence principle are utilized to derive angle-dependent reflection and transmission coefficients, ensuring realistic characterization of RIS performance. Second, a unified EM response model is constructed to bridge Tx-RIS and RIS-Rx sub-channels, enabling consistent and accurate multi-type RIS simulations. Finally, passive and lossless constraints are established to facilitate fair performance comparisons across different RIS configurations.

Simulation results validate the accuracy and practicality of the proposed framework in capturing RIS channel behaviors. The findings highlight that dual-polarized independent control and polarization-rotating RIS offer significant performance advantages in specific scenarios, such as enhanced beamforming capabilities and reduced sidelobe interference, compared to dual-polarized unified control RIS. Furthermore, the angle-dependent modeling reveals critical deviations from idealized performance expectations, emphasizing the importance of precise EM response modeling.

In future work, we aim to validate the proposed framework in various real-world scenarios experimentally. This includes deploying the framework in controlled test environments, such as indoor, outdoor, and high-mobility setups, to assess its accuracy and robustness in characterizing RIS channel behaviors.

\appendices
\section{The unexpanded expression in \eqref{eq34}}
The unexpanded expression in \eqref{eq34} is expanded as follows:
\begin{equation}\label{fundamental_diode_voltage}
\centering
\begin{split}
      {D}_{\phi,0} = &-(4cos\theta_{\mathrm{in}}+2Y_{xx}\eta+2Y_{xx}\eta cos^{2}\theta_{\mathrm{in}}\\&+Y_{xx}^{2}\eta^{2}cos\theta_{\mathrm{in}}+Y_{xy}^{2}\eta^{2}cos\theta_{\mathrm{in}})(4\eta^{2}cos\theta_{\mathrm{in}}\\&+2Z_{xx}\eta+Z_{xx}^{2}cos\theta_{\mathrm{in}}+Z_{xy}^{2}cos\theta_{\mathrm{in}}\\&+2Z_{xx}\eta cos^{2}\theta_{\mathrm{in}}),
\end{split}
\end{equation}
\begin{equation}\label{fundamental_diode_voltage}
\centering
\begin{split}
      {D}_{\phi,1} = &2(2Z_{xx}^{2}cos^{2}\theta_{\mathrm{in}}+2Z_{xy}^{2}cos^{2}\theta_{\mathrm{in}}
    \\&-2Y_{xx}^{2}\eta^{4}cos^{2}\theta_{\mathrm{in}}-2Y_{xy}^{2}\eta^{4}cos^{2}\theta_{\mathrm{in}}\\
    &-4Y_{xx}\eta^{3} cos\theta_{\mathrm{in}}+4Z_{xx}\eta cos^{3}\theta_{\mathrm{in}}\\
    &-2Y_{xx}Z_{xx}\eta^{2}+Y_{xx}Z_{xx}^{2}\eta cos^{3}  \theta_{\mathrm{in}}\\
      &-Y_{xx}^{2}Z_{xx}\eta^{3}cos\theta_{\mathrm{in}}+2Y_{xx}Z_{xx}^{2}\eta^{2} cos^{4}\theta_{\mathrm{in}}\\
      &+Y_{xx}Z_{xy}^{2}\eta cos^{3}\theta_{\mathrm{in}}-Y_{xy}^{2}Z_{xx}\eta^{3}cos\theta_{\mathrm{in}}),
\end{split}
\end{equation}
\begin{equation}\label{fundamental_diode_voltage}
\centering
\begin{split}
      {D}_{\phi,2} = &2\eta cos\theta_{\mathrm{in}}(4Z_{{xy}}cos\theta_{\mathrm{in}}-Y_{xy}Z_{xx}^{2}cos\theta_{\mathrm{in}}\\
      &-Y_{xy}Z_{xy}^{2}cos\theta_{\mathrm{in}}-4Y_{xy}\eta^{2}cos\theta_{\mathrm{in}}) \\
      &+2Y_{xx}Z_{xy}\eta - 2Y_{xy}Z_{xx}\eta\\
      &+Y_{xx}^{2}Z_{xy}\eta ^{2}cos\theta_{\mathrm{in}}+Y_{xy}^{2}Z_{xy}\eta ^{2} cos\theta_{\mathrm{in}}\\&+2Y_{xx}Z_{xy}\eta cos^{2}\theta_{\mathrm{in}}-2Y_{xy}Z_{xx}\eta cos^{2}\theta_{\mathrm{in}}),
\end{split}
\end{equation}
\begin{equation}\label{fundamental_diode_voltage}
\centering
\begin{split}
      {D}_{\phi,3} = &-\eta cos\theta_{\mathrm{in}}(2Y_{xx}Z_{xx}^{2}-8Z_{xx}\\
    &+2Y_{xx}Z_{xy}^{2}-16\eta cos\theta_{\mathrm{in}}\\
    &-8Y_{xx}\eta^{2}cos^{2}\theta_{\mathrm{in}}+Y_{xx}^{2}Z_{xx}^{2}\eta cos\theta_{\mathrm{in}}\\
    &+Y_{xx}^{2}Z_{xy}^{2}\eta cos\theta_{\mathrm{in}}+Y_{xy}^{2}Z_{xx}^{2}\eta cos\theta_{\mathrm{in}}\\
    &+Y_{xy}^{2}Z_{xy}^{2} \eta cos\theta + 2Y_{xx}^{2}Z_{xx} \eta^{2} cos^{2}\theta_{\mathrm{in}}\\
    &+2Y_{xy}^{2}Z_{xx}\eta^{2}cos^{2}\theta_{\mathrm{in}}),
\end{split}
\end{equation}
\begin{equation}\label{fundamental_diode_voltage}
\centering
\begin{split}
      {D}_{\phi,4} = &2\eta cos\theta_{\mathrm{in}}(4Z_{xy}cos\theta_{\mathrm{in}}+Y_{xy}Z_{xx}^{2}cos\theta_{\mathrm{in}}\\
&+Y_{xy}Z_{xy}^{2}cos\theta_{\mathrm{in}}+4Y_{xy}\eta^{2}cos\theta_{\mathrm{in}}\\&+2Y_{xx}Z_{xy}\eta+2Y_{xy}Z_{xx}\eta\\&+Y_{xx}^{2}Z_{xy}\eta^{2}cos\theta_{\mathrm{in}}+Y_{xy}^{2}Z_{xy}\eta^{2}cos\theta_{\mathrm{in}}
      \\&+2Y_{xx}Z_{xy}\eta cos^{2}\theta_{\mathrm{in}}+2Y_{xy}Z_{xx}\eta cos^{2}\theta_{\mathrm{in}}).
\end{split}
\end{equation}
\footnotesize
\bibliographystyle{IEEEtran}
\bibliography{IEEEabrv,Bibliography}
\end{document}